\documentclass[aps, 10pt, prd,
               notitlepage, twocolumn, superscriptaddress,
               longbibliography,
               nofootinbib, floatfix]{revtex4-2}

\usepackage{amsmath}
\usepackage{amssymb}
\usepackage{amsfonts}
\usepackage[utf8]{inputenc}
\usepackage[T1]{fontenc}
\usepackage{mathrsfs}
\usepackage{anyfontsize}
\usepackage{microtype}
\usepackage{eucal}
\usepackage{colortbl}
\usepackage{tabularx}

\usepackage[linktocpage,breaklinks]{hyperref}
\usepackage[usenames,dvipsnames]{xcolor}

\usepackage{tensor}
\usepackage{bm}

\usepackage{graphicx}
\usepackage{epsfig}
\usepackage{epstopdf}

\usepackage{natbib}
\usepackage{tcolorbox}
\usepackage[normalem]{ulem}

\hypersetup{colorlinks=true,
            citecolor=NavyBlue,
            linkcolor=NavyBlue,
            urlcolor=NavyBlue}

\usepackage{tikz}
\usepackage{floatrow}

\newcommand{\real}{\textrm{Re}\,}
\newcommand{\imag}{\textrm{Im}\,}
\newcommand{\mr}[1]{\frac{M^{#1}}{r^{#1}}}
\newcommand{\barm}{\bar{M}}
\newcommand{\css}{c_\textrm{s}^2}
\newcommand{\rh}{r_\textrm{h}}
\newcommand{\ps}{\,(+)}
\newcommand{\mn}{\,(-)}
\newcommand{\nn}{\nonumber \\}
\newcommand{\dd}{{\rm d}}
\newcommand{\ee}{{\rm e}}

\newcommand{\ii}{{\rm i}}
\newcommand{\LL}{{\rm L}}
\newcommand{\RR}{{\rm R}}

\newcommand{\lw}{\ell \omega}

\newcommand{\cA}{{\cal A}}
\newcommand{\cB}{{\cal B}}

\newcommand{\mode}{X^{\, (\pm)}_{\lw}}

\newcommand{\supIn}{\, (\pm),\, {\rm in}}
\newcommand{\supOut}{\, (\pm),\, {\rm out}}

\newcommand{\pmlw}{^{(\pm)}_{\lw}}

\newcommand{\dV}{{\rm d}^{4}x \, \sqrt{-g}}
\newcommand{\lamo}{\lambda_{\rm o}}
\newcommand{\lame}{\lambda_{\rm e}}
\newcommand{\lameo}{\lambda_{\rm e,\, o}}

\newcommand{\Pt}{\tilde{P}}

\newcommand{\mm}{{\rm m}}

\newcommand{\aei}{\affiliation{Max Planck Institute for Gravitational Physics (Albert Einstein Institute),
D-14476 Potsdam, Germany}}
\newcommand{\um}{\affiliation{Departamento de F\'isica, Universidad de Murcia, Murcia, E-30100, Spain}}
\newcommand{\utub}{\affiliation{Theoretical Astrophysics, University of T\"ubingen,
Auf der Morgenstelle 10, D-72076 T\"ubingen, Germany}}
\newcommand{\uv}{\affiliation{Department of Physics, University of Virginia, Charlottesville, Virginia 22904, USA}}
\newcommand{\ethz}{\affiliation{Institut f\"ur Theoretische Physik, ETH Z\"urich, 8093 Z\"urich, Switzerland}}

\begin{document}

\title{Quasinormal modes and their excitation beyond general relativity}

\begin{abstract}
The response of black holes to small perturbations is known to
be partially described by a superposition of quasinormal modes.
Despite their importance to enable strong-field tests of gravity, little to
nothing is known about what overtones and quasinormal-mode amplitudes are like
for black holes in extensions to general relativity.
We take a first step in this direction
and study what is arguably the
simplest  model
that allows first-principle calculations to be made: a
nonrotating black hole in an effective-field-theory extension of general
relativity with cubic-in-curvature terms.
Using a phase-amplitude scheme that uses
analytical continuation and the
Pr\"ufer transformation, we numerically compute, for the first time, the quasinormal
overtone frequencies (in this theory) and quasinormal-mode excitation factors (in any
theory beyond general relativity).
We find that the overtone quasinormal frequencies and their excitation factors
are more sensitive than the fundamental mode to the lengthscale $l$ introduced by the
higher-derivative terms in the effective field theory.
We argue that a description of all overtones cannot be made within the regime of validity of the effective field theory,
and we conjecture that this is a general feature of any extension to general relativity that introduces a new lengthscale.
We also find that a parametrization of the modifications to the general-relativistic quasinormal frequencies
in terms of the
ratio between $l$ and the black hole's mass is somewhat inadequate, and we
propose a better alternative.
As an application, we perform a preliminary study of the implications of the breakdown,
in the effective field theory, of
the equivalence between the quasinormal mode spectra associated to metric perturbations of polar and axial parity of the Schwarzschild black hole
in general relativity.
We also present a simple justification for the loss of isospectrality.
\end{abstract}

\author{Hector O. Silva}    \aei
\author{Giovanni Tambalo}   \ethz \aei
\author{Kostas Glampedakis} \um  \utub
\author{Kent Yagi}          \uv
\author{Jan Steinhoff}      \aei

\maketitle

\section{Introduction}
\label{sec:intro}

The study of the response of black holes to external perturbations has a
long history that dates back to the seminal work of Regge and Wheeler on the
stability of the ``Schwarzschild singularity'' against linear
perturbations in the 1950s~\cite{Regge:1957td}.
Numerical scattering experiments performed by
Vishveshwara~\cite{Vishveshwara:1970cc} showed that the response in time of the
black hole to incident Gaussian wave packets exhibits, after an initial prompt
response, a characteristic damped oscillation, i.e., a ``ringdown.''
The ringdown was also observed in the gravitational-waves signals produced by test
particles plunging radially into black holes by Davis et
al.~\cite{Davis:1971pa,Davis:1971gg,Davis:1972ud} and in stellar collapse by
Cunningham, Price and Moncrief~\cite{Cunningham:1978zfa,Cunningham:1979px}.
The latter showed that the ringdown has oscillation frequency and damping time
in conformity with the quasinormal frequencies of a Schwarzschild black hole
calculated by Chandrasekhar and Detweiler~\cite{Chandrasekhar:1975zza}.
Chandrasekhar and Detweiler also proved a remarkable result: the
spectrum of quasinormal modes associated to metric perturbations of axial parity
(described by Regge-Wheeler equation~\cite{Regge:1957td}) and polar
parity (described by the Zerilli equation~\cite{Zerilli:1970se,Zerilli:1970wzz}) are the
same despite the different forms of these equations. The two spectra are said
to be isospectral.

It was not until the work of Leaver~\cite{Leaver:1986gd} in the 1980s, that the relation between
the source of disturbance and resulting gravitational-wave signal was
studied analytically as an initial-value problem
using Green's functions; see also, e.g., Refs.~\cite{Sun:1988tz,Sun:1990pi,Andersson:1995zk,Nollert:1998ys,Glampedakis:2001js,Berti:2006wq,Zhang:2013ksa}.
It became understood that the ringdown dominates the black hole's response, except at very
early (the ``prompt response'') and very late times (the ``tail''~\cite{Price:1971fb}), and
that it consists of a superposition of quasinormal modes.
The amplitude with which each mode contributes to the ringdown is
determined by its excitation coefficient, which can be factorized into
perturbation independent (termed the ``quasinormal-mode excitation factor'') and dependent parts.
Together, quasinormal modes and their excitation coefficients can be
used to construct the quasinormal-mode contribution to the Green's function
which, in turn, can be used to evolve an initial data in time. This approach
has been used to reproduce the ringdown in the aforementioned cases~\cite{Leaver:1986gd,Andersson:1995zk}.

The quasinormal frequencies of the astrophysically relevant Kerr solution are uniquely determined by the black hole's mass and
spin~\cite{Leaver:1985ax}. The identification of two or more
quasinormal frequencies from gravitational waves produced, e.g., in the coalescence of
binary black holes~\cite{Pretorius:2005gq,Campanelli:2005dd,Baker:2005vv,Buonanno:2006ui},
would enable ``direct evidence of black holes with the same certainty as, say, the
21~cm line identifies interstellar hydrogen,'' as suggested by Detweiler~\cite{Detweiler:1980gk}.
If black hole spectroscopy ever reveals a tension between general
relativistic predictions and observations, it would be suggestive of new physics
beyond general relativity~\cite{Dreyer:2003bv,Berti:2005ys}.

With the advent of gravitational-wave astronomy~\cite{LIGOScientific:2016aoc,LIGOScientific:2016vlm,LIGOScientific:2016lio}, it becomes sensible
to attempt to understand the quasinormal mode spectra (and respective excitation) in theories
beyond general relativity. In general relativity, an ab initio description
of the ringdown for the astrophysically relevant case of comparable mass binary-black hole
coalescence remains an outstanding open problem.
However, as exemplified in the foregoing discussion, progress is possible
within black hole perturbation theory.

Here we take an initial step in this direction. We study the quasinormal mode
spectrum and the excitation factors of a nonrotating black hole in an effective
field theory (EFT) of general relativity~\cite{Endlich:2017tqa}; see the requirements for the construction of the EFT therein.
The motivations behind this choice are manifold.
First, the EFT has only two degrees of freedom, and we avoid
unnecessary technical complications introduced by couplings between the metric
and extra fields, such as scalars.
Second, the EFT admits an exact analytical nonrotating black hole solution to
which perturbation theory can be applied~\cite{Cardoso:2018ptl,deRham:2020ejn,Cano:2021myl,Cano:2023tmv,Cano:2023jbk}.
Third, previous analysis, that focused only on the lowest damped quasinormal
frequencies, showed that isospectrality of perturbations of the Schwarzschild
solution~\cite{Chandrasekhar:1975zza} is broken due to the higher-dimension EFT
terms~\cite{Cardoso:2018ptl,deRham:2020ejn,Cano:2021myl}.
For these reasons, the EFT of general relativity is ideal to address from first principles the following questions,
representative of what may be asked in any extension to general relativity,
\begin{itemize}
    \item \emph{What is the consequence of the breakdown of isospectrality in realistic
            sources of gravitational radiation?}
    \item \emph{How sensitive are quasinormal overtone frequencies to corrections to general relativity?}
\end{itemize}
We give preliminary answers to these questions here.

This work is organized as follows.
In Sec.~\ref{sec:eft_overview}, we review the EFT of general relativity and the black hole
solution we will study.
In Sec.~\ref{sec:bhpt}, we present the equations that describe the linear
perturbations of this black hole. We compare our formulation of the equations
with previous literature, and present a simple explanation for the absence
of isospectrality in the EFT.
In Sec.~\ref{sec:prufer}, we review a phase-amplitude method, developed by one
of us~\cite{Glampedakis:2003dn}, that we use to compute the quasinormal
frequencies (including overtones) and their respective excitation factors.
In Sec.~\ref{sec:results}, we present our numerical results and discuss their regime of validity.
In Sec.~\ref{sec:conclusions}, we summarize and discuss our main results.
We use the mostly plus metric signature and use geometrical units $c = G = 1$.
Parenthesis are used to indicate index symmetrization, as in $T_{(\mu\nu)} =
(T_{\mu\nu} + T_{\nu\mu})/2$.

\section{Effective-field-theory of general relativity}
\label{sec:eft_overview}

\subsection{Action and field equations}
\label{sec:eft_gr}

The general structure of the action is
\begin{equation}
    S = \frac{1}{16 \pi} \int \dV~R
    + \frac{1}{16 \pi} \sum_{n \geqslant 2} {l^{2n - 2}} \, S^{(2n)},
    \label{eq:action_schematic}
\end{equation}
where $l$ is a lengthscale assumed to be small compared to the lengthscale associated with
a black hole of mass $M$, i.e., $M \gg l$, and $S^{(2n)}$ is the action of the $n$th order curvature term
which has $2n$ derivatives of the metric.
For this reason we will use the terminology ``dimension-$2n$ operator.''
Notice that only even powers in $l$ are allowed from dimensional analysis.

One can show that, upon field redefinitions and as long as the EFT construction is built around vacuum GR, that
no dimension-four operators exist. The first nontrivial contribution occurs at dimension six
and, at this order, there are only two operators~\cite{Cano:2019ore}.
The resulting action is
\begin{align}
S &= \frac{1}{16 \pi} \int \dV \,
[R + {l^{4}}  \mathscr{L}],
\label{eq:s6_action_final_cano}
\end{align}
where
\begin{equation}
    \mathscr{L} = \lambda_{\rm e} \, R_{\mu\nu}{}^{\rho\sigma} R_{\rho\sigma}{}^{\delta\gamma}R_{\delta\gamma}{}^{\mu\nu}
    + \lambda_{\rm o} \, R_{\mu\nu}{}^{\rho\sigma} R_{\rho\sigma}{}^{\delta\gamma} \tilde{R}_{\delta\gamma}{}^{\mu\nu} \,,
    \label{eq:l6}
\end{equation}
and
$\tilde{R}_{\mu\nu\rho\sigma} = (1/2) \epsilon_{\mu\nu}{}^{\alpha\beta} R_{\alpha\beta\rho\sigma}$,
where $\epsilon_{\mu\nu\rho\sigma}$ is the totally antisymmetric Levi-Civita tensor,
$\lameo$ are dimensionless constants
associated to the even- (``e'') and odd-parity (``o'') curvature terms, and
$\ell$ is a lengthscale.

The field equations of the theory, obtained by varying the action~\eqref{eq:s6_action_final_cano} with respect to $g^{\mu\nu}$, are
\begin{equation}
    \mathscr{E}_{\alpha\beta} = G_{\alpha\beta}
    + l^{4} \mathscr{S}_{\alpha\beta} = 0,
    \label{eq:field_equations_schematic}
\end{equation}
where
\begin{subequations}
\label{eq:field_equations}
\begin{align}
    \mathscr{S}_{\alpha\beta}
    &=
    P_{(\alpha}{}^{\rho\sigma\gamma} R_{\beta)\rho\sigma\gamma}
    - \tfrac{1}{2} g_{\alpha\beta} \mathscr{L}
    + 2 \nabla^{\sigma} \nabla^{\rho} P_{(\alpha|\sigma|\beta)\rho},
    \nonumber \\
    \label{eq:def_field_eqs}
    \\
    P_{\alpha\beta\mu\nu}
    &=
    3 \lame R_{\alpha\beta}{}^{\rho\sigma} \, R_{\rho\sigma\mu\nu}
    \nonumber \\
    &\quad + \tfrac{3}{2} \lamo (R_{\alpha\beta}{}^{\rho\sigma} \tilde{R}_{\rho\sigma\mu\nu}
    + R_{\alpha\beta}{}^{\rho\sigma} \tilde{R}_{\mu\nu\rho\sigma}).
    \label{eq:def_p_tensor}
\end{align}
\end{subequations}
We will only consider the even-parity operator hereafter, i.e., we set $\lamo = 0$.
For brevity, we will write $\lame = \lambda$, and assume $\lambda$ to be positive.
A priori, however, $\lambda$ can have either sign; see, e.g.,~Refs.~\cite{Goon:2016mil,Caron-Huot:2022ugt,Horowitz:2023xyl}.
We will work to leading order in $\lambda$, that is, to $\mathcal{O}(l^4)$.
Other aspects of the EFT in the context of gravitational-wave
physics are discussed, e.g., in Refs.~\cite{Sennett:2019bpc,AccettulliHuber:2020dal,Silva:2022srr,Cano:2022wwo,Cayuso:2023xbc,Melville:2024zjq} and references therein.

\subsection{Nonrotating black hole solution}
\label{sec:bh_solution}

A nonrotating spherically symmetric black-hole solution of the
theory~\eqref{eq:s6_action_final_cano} was found in Refs.~\cite{Cano:2019ore,deRham:2020ejn,Cano:2021myl}.
The line element is
\begin{equation}
    \dd s^2 =
    - N^2 f \, \dd t^2
    + f^{-1} \, \dd r^2
    + r^2 \, \dd \theta^2
    + r^2 \sin^2\theta \, \dd\varphi^2,
    \label{eq:line_element}
\end{equation}
where the metric functions $N$ and $f$ are, respectively,
\begin{subequations}
\label{eq:metric_Nf}
\begin{align}
    N &= 1 - 108 \, \varepsilon \frac{M^6}{r^6} \,,
    \label{eq:metric_N}
    \\
    f &= 1 - \frac{2M}{r}
    + 216 \, \varepsilon \left( 1 - \frac{49}{27} \frac{M}{r} \right) \frac{M^6}{r^6} ,
    \label{eq:metric_f}
\end{align}
\end{subequations}
and we introduced the dimensionless parameter,
\begin{equation}
\varepsilon = {\lambda \, l^{4}} / {M^4} \,,
\end{equation}
and $M$ is the Arnowitt-Deser-Misner (ADM) mass of the black hole.
The event horizon $\rh$ is located at the largest positive root of $f$,
\begin{equation}
    \rh = 2 M ( 1 - 5 \, \varepsilon / 16 ).
    \label{eq:event_horizon}
\end{equation}
The black-hole solution reduces to that of Schwarzschild in the limit $\lambda \to 0$, i.e., $\varepsilon \to 0$. We note that $\rh$ can vanish if $\varepsilon = 16/5$ which, however, is outside the validity of the EFT, $\varepsilon \ll 1$.
For the sake of completeness, we derive
Eq.~\eqref{eq:line_element} in Appendix~\ref{app:bh_derivation}. Further,
in Appendix~\ref{app:petrov}, we show that the spacetime is of Petrov-type D.

The spacetime as written in Eq.~\eqref{eq:line_element} demands some care when
$r$ is approximately $2M$. To see this, take the $g_{rr}$ metric component,
\begin{equation}
f^{-1} =
\left[ 1 - 2M/r + \varepsilon \, \delta f(r) \right]^{-1} \,,
\label{eq:grr_not_resum}
\end{equation}
where we defined
\begin{equation}
\delta f = 216 \, \frac{M^6}{r^6} \left( 1 - \frac{49}{27} \frac{M}{r} \right).
\label{eq:delta_f}
\end{equation}
As $r \to 2M$, we see that the EFT correction starts dominating over the general-relativistic term.
This means that the expansion in $\varepsilon$ ceases to hold in this limit. The same happens for $g_{tt} = - N^2 f$.

To resolve this issue we perform a ``resummation.''
The idea is to factor out a multiplicative
term $1 - \rh/r$, thereby recasting either $N^2 f$ or $f$
in the schematic form:
\begin{equation}
z(r)  = (1 - \rh/r)~[1 + \varepsilon \, \delta z(r)],
\quad z~\textrm{arbitrary}.
\end{equation}
In this way, we guarantee that ${\cal O}(\varepsilon)$ terms
are small for any value of $r \geqslant \rh$.
We leave the details of this calculation to Appendix~\ref{app:resummation}
and quote our final result:
\begin{subequations} \label{eq:resum_gtt_grr}
    \begin{align}
    \label{eq:gtt_resum_final}
    N^2 f &=
    \left(1 - \frac{\rh}{r}\right)\left[ 1 - \varepsilon \, \left(
    \frac{5 M}{8 r}+\frac{5 M^2}{4 r^2}+\frac{5 M^3}{2 r^3}+\frac{5 M^4}{r^4}
    \right. \right.
    \nonumber \\
    &\quad \left.\left. +\frac{10 M^5}{r^5}+\frac{20 M^6}{r^6}\right)\right],
    \\
    \label{eq:resum_f}
    f^{-1} &= \left(1-\frac{\rh}{r}\right)^{-1}  \, \left[
    1 + \varepsilon
    \left( \frac{5 M}{8 r} + \frac{5 M^2}{4 r^2} + \frac{5 M^3}{2 r^3} + \frac{5 M^4}{r^4}
    \right.
    \right.
    \nn
    & \left.\left. \quad
    + \frac{10 M^5}{r^5} - \frac{196 M^6}{r^6}
    \right)\right].
    \end{align}
\end{subequations}
Equations~\eqref{eq:gtt_resum_final} and~\eqref{eq:resum_f} are the expressions
we will use for the $g_{tt}$ and $g_{rr}$ metric components, respectively.
By construction, both equations have the expected behavior at the event horizon $\rh$.
In addition, since to ${\cal O}(\varepsilon)$,
\begin{equation}
N^2 f = f \simeq 1 - 2 M / r  + {\cal O}(r^{-2}), \quad r/M \gg 1,
\end{equation}
we have retained the interpretation of $M$ being the ADM black-hole mass
and that the spacetime is asymptotically Minkowski.

\section{Black hole perturbations}
\label{sec:bhpt}

The linear gravitational perturbations of the black-hole solution~\eqref{eq:line_element} were analyzed in
Refs.~\cite{deRham:2020ejn,Cano:2021myl}, in the Regge-Wheeler-Zerilli formalism (``metric-perturbation approach'')~\cite{Regge:1957td,Zerilli:1970se,Zerilli:1970wzz}, and in Refs.~\cite{Cano:2023tmv,Cano:2023jbk}
in the Newman-Penrose~\cite{Newman:1961qr} and Geroch-Held-Penrose~\cite{Geroch:1973am} formalisms (``curvature-perturbation approach''). See also Refs.~\cite{Hussain:2022ins,Li:2022pcy} for related work in the latter approach.

In the metric-perturbation approach, the problem reduces to studying two
equations in the frequency domain
\begin{equation}
    \left[ \frac{\dd^2}{\dd x^2} + \frac{\omega^2}{\css(r)} - V^{(\pm)}_{\ell}(r) \right]
    X^{(\pm)}_{\lw}(r) =
    0,
    \label{eq:eqs_rwz_fd}
\end{equation}
that we now describe in detail.
The superscript $(\pm)$ is used to denote variables associated to metric
perturbations of polar ($+$) or axial ($-$) parity, which we assume
to have harmonic time-dependence $\exp( - \ii \omega t )$, and are labeled
by the multipole index $\ell \geqslant 2$.
Metric perturbations of polar and axial parities are fully described by a single master function known
as the Zerilli $X^{(+)}$ and Regge-Wheeler $X^{(-)}$ functions, respectively.\footnote{
We verified that in the EFT, as in general relativity, the Zerilli~\cite{Zerilli:1970se} and Zerilli-Moncrief~\cite{Moncrief:1974am} functions
and the Regge-Wheeler~\cite{Regge:1957td} and Cunningham-Price-Moncrief~\cite{Cunningham:1978zfa} functions satisfy
the same homogeneous differential equations.}
We also introduced the tortoise coordinate $x$, defined as
\begin{equation}
\label{eq:def_tortoise}
    \dd x / \dd r = 1 / (Nf),
\end{equation}
and that maps the domain $\rh \leqslant r < \infty$ to $-\infty < x < \infty$.
This is not guaranteed to happen for all values of $\varepsilon$.
As we detail in Appendix~\ref{app:tortoise}, the desired mapping
holds for $\varepsilon \lesssim 0.59$.
Finally, $V_{\ell}^{\,(\pm)}$ and $c_{\rm s}^2$ are the black-hole effective potential
and propagation velocity of the perturbations, respectively. Let us consider
both in turn.

\begin{figure*}[t]
\includegraphics{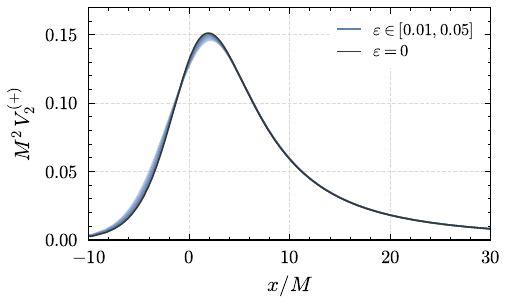}
\includegraphics{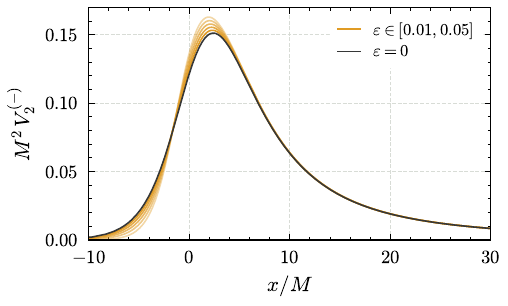}
\caption{The effective potentials $V_{2}^{(\pm)}$ for perturbations of polar (left panel) and axial (right panel) parity. We vary the parameter $\varepsilon = \lambda \, l^4/M^4$ from zero (general relativity) to $0.05$
in increments of $\Delta \varepsilon = 0.01$.
The value of the potential's peak decreases (increases) for the polar- (axial-) parity potentials with
respect to general relativity. The location of the peak shifts in opposite directions, with respect to the case of general relativity: outwardly for polar-parity
and inwardly for the axial-parity potential. These changes are bound to the region between the event horizon, pushed to $x \to -\infty$, and the location of the potential peak.}
\label{fig:potentials}
\end{figure*}

First, the effective potential can be decomposed in a resummed form
as
\begin{equation}
    V_{\ell}^{\,(\pm)} = \left(1 - \frac{\rh}{r} \right) \,
    \left[ \bar{V}_{\ell}^{\,(\pm)} + \varepsilon \, \delta V_{\ell}^{\,(\pm)} \right].
\end{equation}
The bare general-relativity contributions to the potential are the Zerilli~\cite{Zerilli:1970se} and Regge-Wheeler~\cite{Regge:1957td} potentials, given respectively by
\begin{subequations}
\label{eq:effective_potentials_gr}
\begin{align}
    \bar{V}_{\ell}^{\ps} &= \frac{1}{(r \Lambda_{\ell})^2} \left[
        2 \lambda_{\ell}^2 \left(\Lambda_{\ell} + 1\right)
        + \frac{18 M^2}{r^2} \left( \lambda_{\ell} + \frac{M}{r} \right)
    \right],
    \nonumber \\
    \label{eq:pot_zerilli}
    \\
    \bar{V}_{\ell}^{\mn} &= \frac{1}{r^2} \left[ \ell(\ell+1) - \frac{6M}{r} \right],
    \label{eq:pot_regge_wheeler}
\end{align}
\end{subequations}
where we defined
\begin{equation}
    \lambda_{\ell} = (\ell + 2) (\ell - 1) / 2,
    \quad \textrm{and} \quad
    \Lambda_{\ell} = \lambda_{\ell} + 3M/r.
\label{eq:def_lambdas}
\end{equation}
The modifications to these potentials originating from the dimension-six operators can be written schematically as
\begin{subequations}
\label{eq:effective_potentials_eft_schematic}
\begin{align}
    \delta V_{\ell}^{\ps} &= \frac{1}{(r \Lambda_{\ell})^2} \, \sum_{i=1}^{10} v^{\ps}_{i \ell}(r) \, (M/r)^i ,
    \\
    \delta V_{\ell}^{\mn} &= \frac{1}{r^2} \, \sum_{i=1}^{7} v^{\mn}_{i \ell} \, (M/r)^i .
\end{align}
\end{subequations}
The coefficients
$v^{\ps}_{i \ell}$ contain $\Lambda_{\ell}$ for $n>4$; hence the explicitly stated
dependence on $r$.
In contrast, $v^{\mn}_{i \ell}$ is independent of $r$ for all $n$.
All coefficients have powers of $\ell$.
We show the explicit forms of $v^{\,(\pm)}_{i \ell}$ in Appendix~\ref{app:effective_potential_coeffs}.

In Fig.~\ref{fig:potentials} we show both potentials, Eqs.~\eqref{eq:pot_zerilli} and~\eqref{eq:pot_regge_wheeler},
for $\ell = 2$ as functions of the tortoise coordinate $x$.
The curves correspond to increasing values of $\varepsilon$,
from zero (general relativity) to $0.05$ in steps of $\Delta \varepsilon = 0.01$.
The EFT corrections are most salient in the region between the event horizon
and the location of the potential peak; past the latter the curves become
identical to one another.

It is important to note that the potentials are short ranged,
i.e., their integral on the domain $x \in (-\infty,\,+\infty)$
is finite. Indeed, one can verify that
\begin{equation}
    \int_{-\infty}^{+\infty}
    V_{\ell}^{\,(\pm)} \, \dd x
    = \frac{1}{2M} \left[2 \lambda_{\ell} + \frac{1}{2}
    + \varepsilon \, \iota_{\ell}^{\,(\pm)} \right],
    \label{eq:miura}
\end{equation}
where $\iota_{\ell}^{\ps} \neq \iota_{\ell}^{\mn}$ are parity-dependent functions of $\ell$.
In the limit of general relativity, the integrals of the Regge-Wheeler and
Zerilli potentials are the same.
This equality, first noticed by Chandrasekhar and Detweiler~\cite{Chandrasekhar:1975zza} (see also Refs.~\cite{1980RSPSA.369..425C,Chandra:SchiffLectures}),
is a necessary condition to establish the isospectrality of
the Regge-Wheeler and Zerilli potentials. More precisely, the equality of Eq.~\eqref{eq:miura} is
the first of an infinite hierarchy of integral equalities that must be satisfied by a
pair of potentials if they are to have the same reflection and transmission coefficients.\footnote{These integrals
are formally related to conserved quantities allowed by the Korteweg-de Vries equation~\cite{Miura:1968JMP.....9.1204M}.
See Ref.~\cite{1980RSPSA.369..425C}, Sec.~4 and its Appendix, for a discussion, and Refs.~\cite{Glampedakis:2017rar,Lenzi:2021njy} for recent literature.}
That this equality is broken by the EFT corrections already at ``zeroth order'' in this hierarchy implies the breakdown of isospectrality.

Finally, perturbations of polar and axial parity propagate with a position-dependent velocity
\begin{align}
    \css &= 1 - 288 \, \varepsilon \, f \, \frac{M^5}{r^5},
    \nn
    &\simeq 1 - 288 \, \varepsilon \left( 1 - \frac{\rh}{r} \right) \frac{M^5}{r^5} \,,
    \label{eq:sounds_speed}
\end{align}
where in the second line we used the resummed form of $f$ [cf.~Eq.~\eqref{eq:resum_f}] and kept the ${\cal O}(\varepsilon)$ term only. We may note that $\css$ is unity at spatial infinity and at the event horizon $\rh$, while it can be
sub- or superluminal outside $\rh$ depending on the sign of $\varepsilon$.
Reference~\cite{deRham:2020zyh} argues that this
speed cannot be used to predict time delay (or advance) with respect
to general relativity as long as one stays within the regime
of validity of the EFT.

Before proceeding, let us compare our Eq.~\eqref{eq:eqs_rwz_fd} with those
found in the literature, particularly in the works by de Rham et
al.~\cite{deRham:2020ejn} and Cano et al.~\cite{Cano:2021myl} who, like ourselves,
worked in metric perturbation approach.
In comparison to Ref.~\cite{deRham:2020ejn}, our perturbation equations are
similar to their Eqs.~(2.23) except that we did a resummation of the
effective potentials. Likewise, we use the same definition of the tortoise
coordinate, though, again, we perform a resummation. To compute the quasinormal
mode frequencies, Ref.~\cite{deRham:2020ejn} recasts their equations in a form
that can be mapped into the quasinormal frequency parametrization of Ref.~\cite{Cardoso:2019mqo}.
In comparison to Ref.~\cite{Cano:2021myl}, our perturbation equations are
different both in the choice of the tortoise coordinate [they use
a ``pseudotortoise coordinate''; see Eq.~(41) therein] and in the effective potentials
(they do a field redefinition to trade the position-dependent
propagation speed $c^2_{\rm s}$ for a frequency-dependent potential).
To compute the
quasinormal frequencies, Ref.~\cite{Cano:2021myl} did a direct integration
of the perturbation equations.\footnote{Implicitly, Ref.~\cite{deRham:2020ejn}
also does a direct integration of the perturbations equations. The reason is that the
theory-agnostic coefficients in Ref.~\cite{Cardoso:2019mqo} are
found by direct integration of their parametrized perturbation equation; see Eq.~(10) and Sec.~III therein.}

Cano et al.~\cite{Cano:2021myl} reports an agreement of approximately $1\%$
to $5\%$ of their results with those of de Rham et al.~\cite{deRham:2020ejn}.
In another work, Cano et al.~\cite{Cano:2023jbk} also computed the quasinormal frequencies for rotating black holes using the curvature-perturbation approach and using a small spin expansion~\cite{Cano:2023tmv}. They followed the approach of Ref.~\cite{Hussain:2022ins} to compute the quasinormal frequencies,
and the results were cross-checked against the direct integration of the modified Teukolsky equation. In the nonrotating limit, their results agree with those of Refs.~\cite{Cano:2021myl,deRham:2020ejn}.
Our results will be presented in Sec.~\ref{sec:results}.
However, first, let us motivate and explain the phase-amplitude method we will
adopt to compute the quasinormal modes and their excitation.

\section{Quasinormal modes and their excitation}
\label{sec:prufer}

In this section, we review the ``quick and dirty'' phase-amplitude
method developed by Glampedakis and Andersson~\cite{Glampedakis:2003dn}
for studying black-hole resonances.
We will first review some technical difficulties in numerically computing
quasinormal modes and how they are overcome in the phase-amplitude approach.
We will then explain how the quasinormal-mode excitation factors can be
determined.

\subsection{Quasinormal modes}
\label{sec:qnm_review}

We are interested in computing the quasinormal modes associated
to Eq.~\eqref{eq:eqs_rwz_fd}. Because the effective potentials
are short ranged and vanish both at the horizon and at spatial infinity, whereat the propagation speed
$c^2_{\rm s}$ becomes
unity, the general physical
solution of Eq.~\eqref{eq:eqs_rwz_fd} has the form
\begin{align}
    \mode \simeq
    \begin{cases}
        \, \ee^{- \ii \omega x} \quad &x \to -\infty\\
        \,
        A^{\supIn}_{\lw}    \, \ee^{- \ii \omega x}
        + A^{\supOut}_{\lw} \, \ee^{+ \ii \omega x}
        \quad &x \to +\infty
    \end{cases} \,,
    \label{eq:bcs_general}
\end{align}
consisting of purely ingoing waves at the event horizon and a mixture
of ingoing and outgoing waves at spatial infinity.
From the ratio between the amplitudes of the ingoing and outgoing waves
at spatial infinity, we can define the scattering matrix
\begin{equation}
    \mathcal{S}^{(\pm)}_{\lw} = (-1)^{\ell + 1} A_{\lw}^{\supOut} / A_{\lw}^{\supIn} = \exp[2 \ii \delta^{(\pm)}_{\lw}],
    \label{eq:s_matrix}
\end{equation}
where $\delta^{(\pm)}_{\lw}$ is the phase-shift function.
Quasinormal modes are solutions defined by having $A_{\lw}^{\supIn} = 0$,
i.e., they are the poles of the scattering matrix~\cite{Vishveshwara:1970cc}.
The problem of computing the quasinormal-mode frequencies $\omega_{\ell n}$
hence reduces to a boundary-value problem in which one has to find $\omega_{\ell n}$
such that $A^{\supIn}_{\lw}$ vanishes. Root-finding algorithms can be used to perform this task.
In black-hole physics, for each multipole $\ell$, there is an infinite number $n$ of quasinormal frequencies
that we sort according to their damping time. The index $n=0$ is used for quasinormal-mode frequency
with longest decay time (the ``fundamental mode'') and modes with $n > 0$ are called ``overtones.''

A numerical challenge immediately presents itself if one attempts to
carry such procedure by numerically integrating the differential
equation~\eqref{eq:eqs_rwz_fd}.
Since $\omega_{\ell n}$ is complex-valued, we find
\begin{equation}
    X^{(\pm)}_{\lw} \simeq \exp( \mp \, x \, \imag \omega_{\ell n} ),
    \quad x \to \pm \infty.
\end{equation}
and because $\imag \omega_{\ell n}$ is negative for stable perturbations,
the quasinormal-mode solution diverges as $x \to \pm \infty$.
Consequently, in the root-finding process, we must resolve an exponentially decaying from an exponentially growing part of the solution, at large values of $x$.
This is even more challenging for overtones which, by definition, have shorter damping times.

\subsection{The phase-amplitude method}
\label{sec:phase_amp}

Reference~\cite{Glampedakis:2003dn} proposed a ``quick-and-dirty'' method for the calculation of
the quasinormal-mode frequencies that combines two ideas. First, instead of working
with the (possibly rapidly varying) function $X\pmlw$, one works with slowly varying phase functions.
Second, instead of working on the real axis, one performs an analytical continuation of $X\pmlw$ to complex values of $x$
and a suitable integration path is chosen in order to balance the exponentially decaying and growing waves of the general solution Eq.~\eqref{eq:eqs_rwz_fd}.
Let us see how this works in practice. To lighten the notation, we will omit
the parity ``$(\pm)$'' and mode ``$\lw$'' scripts for now.

We start by rewriting Eq.~\eqref{eq:eqs_rwz_fd} as
\begin{equation}
\left[ \frac{\dd^2}{\dd x^2} + Q \right] X = 0,
\quad Q = \omega^2 / c_{\rm s}^2 - V,
\label{eq:general_fd_Q}
\end{equation}
where $Q \to \omega^2$ as $x \to \pm \infty$,
and the boundary conditions~\eqref{eq:bcs_general} as follows:
\begin{equation}
    X \simeq
    \begin{cases}
        \, \ee^{- \ii \omega x} \quad &x \to -\infty\\
        \,
        B \sin( \omega x + \zeta )
        \quad &x \to +\infty
    \end{cases} \,,
    \label{eq:bcs_general_sin}
\end{equation}
where $B$ and $\zeta$ are complex-valued constants.
Equation \eqref{eq:general_fd_Q} admits an exact solution in the form
\begin{equation}
    X = \exp\left[ \int P(x') \, \dd x'\right],
\end{equation}
where $P$ is the \emph{phase function},
\begin{equation}
    P = \dd \log X / \dd x,
\end{equation}
which, from Eq.~\eqref{eq:general_fd_Q}, satisfies the Riccati
equation\footnote{We may, parenthetically here, remark that the advantage of working
with the Riccati equation had already been appreciated by Chandrasekhar and Detweiler~\cite{Chandrasekhar:1975zza}; cf.~pp.~451 therein.}
\begin{equation}
    \dd P / \dd x + P^2 + Q = 0,
    \label{eq:riccati}
\end{equation}
and the quasinormal-mode boundary conditions translate into $P \to - \ii \omega$ as $x \to - \infty$.

Instead of working with $P$ as $x \to \infty$ as well, it is useful to introduce a second phase function, $\Pt$, by means of the \emph{Pr\"ufer transformation} defined as:
\begin{subequations}
\label{eq:prufer_definition}
\begin{align}
X &= B \sin[ \omega x + \Pt(x) ], \label{eq:prufer_X}
\\
\dd X / \dd x &= B \omega \cos[ \omega x + \Pt(x) ].
\label{eq:prufer_dX}
\end{align}
\end{subequations}
We can calculate $\dd \log X / \dd x$ with the foregoing equations
and find that the phase functions $P$ and $\Pt$ are related by
\begin{equation}
    P = \omega \cot( \omega x + \Pt ),
    \label{eq:phase_from_prufer}
\end{equation}
with inverse
\begin{equation}
    \Pt = - \omega x + \frac{1}{2\ii} \log\left[ \frac{\ii P - \omega}{\ii P + \omega} \right].
    \label{eq:prufer}
\end{equation}
Note that Eq.~\eqref{eq:phase_from_prufer} explains the absence of a $\dd \Pt / \dd x$ term in Eq.~\eqref{eq:prufer_dX}.
A short calculation shows that the Pr\"ufer phase function satisfies
\begin{equation}
    \dd \Pt / \dd x + (\omega - Q / \omega) \sin^2(\omega x + \Pt) = 0.
    \label{eq:prufer_eq}
\end{equation}

From the asymptotic properties of $Q$ and $P$, we find from
Eqs.~\eqref{eq:riccati} and~\eqref{eq:prufer_eq} that for real $\omega$ and $\ell$,
\begin{subequations}
\begin{align}
    \dd P / \dd x &\simeq 0, \quad x \to - \infty
    \\
    \dd \Pt / \dd x &\simeq 0, \quad x \to + \infty
\end{align}
\end{subequations}
That is, $P$ and $\Pt$ are slowly varying functions as $x \to -\infty$
and $x \to \infty$, respectively.\footnote{Equations~\eqref{eq:riccati} and~\eqref{eq:prufer_eq}
can be used to compute Regge poles~\cite{Glampedakis:2003dn}. Like quasinormal modes, they are poles
of the scattering matrix~\eqref{eq:s_matrix}, but correspond instead to complex values of $\ell$
for a given real-valued $\omega$. Regge poles are important in scattering theory; see, e.g.,~Refs.~\cite{Andersson:1994rm,Andersson:1994rk,Decanini:2002ha}
for applications in black-hole physics.}
It is then suggestive that we should work with $P$ in the domain $x \in (-\infty, x_\mm]$
and with $\Pt$ in the domain $x \in (x_\mm, \infty)$, where $x_\mm$ is a matching point.
Experience has shown that the computation of the quasinormal frequencies does
not depend on the precise value of $x_\mm$, as long as we chose it to be near the
peak of the effective potential $V$~\cite{Glampedakis:2003dn}.
Here we chose $r_\mm  = 3 M$, the location of the light ring in the
Schwarzschild spacetime, which translates to $x_{\mm} \simeq 1.61 M$.

Finally, we can compare Eqs.~\eqref{eq:prufer_X} and~\eqref{eq:bcs_general_sin}
to conclude that $\Pt \to \zeta$ as $x \to \infty$
and, by comparing Eqs.~\eqref{eq:bcs_general} and~\eqref{eq:bcs_general_sin}, that
\begin{equation}
    A_{\rm in} / A_{\rm out} = - \exp(- 2 \ii \zeta).
    \label{eq:ain_aout_prufer}
\end{equation}

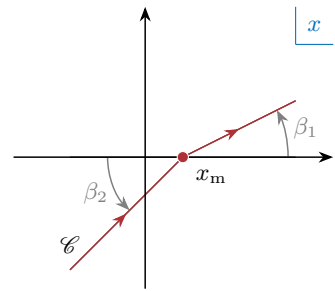
\begin{figure}[t] \label{fig:tidal-graphs2}
\centering
\usetikzlibrary{arrows.meta}
\usetikzlibrary{ decorations.markings}
\usetikzlibrary {angles,quotes}
\begin{tikzpicture}[
    baseline={(current bounding box.center)},
    scale=1,
    >=Stealth,
    decoration={markings, mark=at position 0.5 with {\arrow{Stealth}}}
    ]

    \coordinate (x_plus_inf) at (2.,0.75);
    \coordinate (x_match) at (0.5,0);
    \coordinate (x_minus_inf) at (-1.,-1.5);

    \draw[NavyBlue] (2,1.5) -- (2,2);
    \draw[NavyBlue] (2,1.5) -- (2.5,1.5);
    \coordinate [label=center:$\textcolor{NavyBlue}{x}$] (A) at (2.25, 1.75);

    \coordinate (A) at (2,0);
    \coordinate (B) at (x_match);
    \coordinate (C) at (x_plus_inf);
     \draw (A) -- (B) -- (C)
        pic [->,draw=gray, line width=0.2mm, angle radius=14mm,
            "${\textcolor{gray}{\beta_1}}$", angle eccentricity=1.2] {angle = A--B--C};

    \coordinate (A) at (-1,0);
    \coordinate (B) at (x_match);
    \coordinate (C) at (x_minus_inf);
     \draw (A) -- (B) -- (C)
        pic [->, draw=gray, line width=0.2mm, angle radius=10mm,
        "${\textcolor{gray}{\beta_2}}$", angle eccentricity=1.25] {angle = A--B--C};

    \draw[-,line width=0.2mm,postaction={decorate},Maroon] (x_match) -- (x_plus_inf);
    \draw[-,line width=0.2mm,postaction={decorate},Maroon] (x_minus_inf) -- (x_match);

    \draw[->,line width=0.2mm] (-1.75,0) -- (2.5,0);
    \draw[->,line width=0.2mm] (0,-1.75) -- (0,2);

    \node [fill=Maroon, draw=White, inner sep=1.5pt, circle, label=-45:$x_{\textrm{m}}$] at (x_match) {};
    \node [label=$\mathscr{C}$] at (x_minus_inf) {};

\end{tikzpicture}
\caption{The integration path $\mathscr{C}$ in the complex $x$-plane used for the calculation
of the quasinormal modes. The value of $\beta_{1,2}$ is the chosen based on the asymptotic
behavior of $Q$ as $x \to \pm \infty$. In our problem, $\beta_{1,2} = - \arg \omega$. However, the
method can be applied to other situations where this is not the case, such as of perturbations of the Kerr black hole~\cite{Glampedakis:2003dn}.}
\label{fig:int_cont}
\end{figure}

The second step in the scheme of Ref.~\cite{Glampedakis:2003dn} consists
in promoting the tortoise coordinate $x$ to become complex valued as in, e.g.,
in the closely related phase-integral approach~\cite{Andersson:1992scr}.
Consider the curve $\mathscr{C}$ illustrated in Fig.~\ref{fig:int_cont},
\begin{equation}
    x = x_{\mm} + \rho \exp( \ii \beta),
    \label{eq:complex_tortoise}
\end{equation}
parametrized by the real parameter $\rho \in (-\infty,\,\infty)$
and where $\beta$ is a real constant. Note that the matching point is at $\rho = 0$.
In terms of Eq.~\eqref{eq:complex_tortoise}, the Pr\"ufer
equation~\eqref{eq:prufer_eq}
becomes
\begin{equation}
    \dd \Pt / \dd x + (\omega - Q / \omega) \sin^2 [\omega x_\mm  + \omega \rho \exp(\ii \beta) + \Pt]
    = 0.
    \label{eq:prufer_tmp}
\end{equation}
The crucial step is to now observe that, with a suitable choice of $\beta$, we
can make the ingoing and outgoing waves as $x \to
\infty$ to be of comparable asymptotic amplitudes.
This is achieved by choosing,
\begin{equation}
    \beta = - \arg \omega,
    \label{eq:def_beta}
\end{equation}
such that the integration path is parallel to the anti-Stokes lines when $x \to
\pm \infty$~\cite{Glampedakis:2003dn}. Equation~\eqref{eq:prufer_tmp} becomes
\begin{equation}
    \dd \Pt / \dd x
    + (\omega - {Q}/{\omega})
    \sin^2(\omega x_\mm + |\omega| \rho + \Pt) = 0.
    \label{eq:prufer_eq_beta}
\end{equation}
Note that now the argument term proportional to $\rho$ is real and hence we
eliminated the asymptotic exponential behavior of the amplitude.

We can then rewrite the differential equations~\eqref{eq:riccati} and~\eqref{eq:prufer_eq_beta} for the phase functions
with $\rho$ as an independent variable, and trade $x$ in favor of $r$ by means
of Eq.~\eqref{eq:complex_tortoise}. We obtain
\begin{subequations} \label{eq:final_qnm_eqs}
\begin{align}
    &\frac{\dd P}{\dd \rho} + \ee^{\ii \beta} ( P^2 + Q ) = 0,
    \label{eq:dP_drho}
    \\
    &\frac{\dd \Pt}{\dd \rho} + \ee^{\ii \beta} \left(\omega - \frac{Q}{\omega} \right)
    \sin^2(\omega x_\mm + |\omega| \rho + \Pt) = 0,
    \label{eq:dPt_drho}
    \\
    &\frac{\dd r}{\dd \rho} - \ee^{\ii \beta} \frac{\dd r}{\dd x} = 0,
    \label{eq:dr_drho}
\end{align}
\end{subequations}
where $\dd r / \dd x$ and $\beta$ are given by Eqs.~\eqref{eq:def_tortoise}
and~\eqref{eq:def_beta}, respectively.

Equation~\eqref{eq:final_qnm_eqs} constitutes the system
of differential equations we need to integrate to compute the quasinormal frequencies.
The integration procedure can be summarized as follows:
\begin{enumerate}
    \item Choose the values of $\varepsilon$, $\ell$, and $\omega$, and with the latter
    compute $\beta = - \arg \omega$.
    \label{step:lme}
    \item Determine the initial condition for $r$ as $\rho \to -\infty$.
    We found it useful, particularly in the context of the integrations in the next
    section, to integrate Eq.~\eqref{eq:dr_drho} from $\rho_\mm = 0$
    (at which $r = 3M$) backward to, say,~$\rho_{\rm min} = - 50M$. This fixes
    $r_{\rm min} = r(\rho_{\rm min})$.
    \item Set the initial condition for $P$ at $\rho_{\rm min}$ using
    the leading-order Wentzel-Kramers-Brillouin (WKB) formula, that is,
    $P_{\rm min} = - \ii Q(r_{\rm min})^{1/2}$.
    \item Integrate Eqs.~\eqref{eq:dP_drho} and~\eqref{eq:dr_drho} from $\rho_{\rm min}$ to $\rho_\mm$,
    using $r_{\rm min}$ and $P_{\rm min}$ as initial conditions.
    \item Calculate $\Pt(\rho_\mm)$ from $P(\rho_\mm$), the result of the previous step, using Eq.~\eqref{eq:prufer}.
    \item Switch to Pr\"ufer phase function, that is, integrate Eqs.~\eqref{eq:dPt_drho} and~\eqref{eq:dr_drho} from $\rho = \rho_\mm$ up to $\rho = \rho_{\rm max}$. In practice, we often used $\rho_{\rm max} \approx 10^4 M$. This gives us $\zeta \simeq \Pt(\rho_{\rm max})$.
    \item Calculate $A_{\rm in}/A_{\rm out}$ using Eq.~\eqref{eq:ain_aout_prufer}.
    \label{step:ain_aout_ratio}
    \item Repeat steps~\ref{step:lme} to~\ref{step:ain_aout_ratio}, updating the value of $\omega$
    until the quasinormal-mode boundary condition $A_{\rm in} = 0$ is satisfied. This is a root-finding
    problem that we solve using Muller's method~\cite{Muller:MR0083822}.
\end{enumerate}

We implemented the foregoing steps in C++. The integration of the differential
equations was performed with the Runge-Kutta-Fehlberg (7,8) method, as implemented
in Odeint~\cite{odeint}, part of the Boost library~\cite{Boost}.
Our implementation of Muller's method follows the pseudocode found in the
``Numerical Recipes,'' Chapter~9.2~\cite{Press2007Numerical}.
We will present the numerical results of our quick-and-dirty quasinormal mode
frequency computations in Sec.~\ref{sec:qnm}.

\subsection{The excitation factors}
\label{sec:method_exc_fac}

Having explained how we compute the quasinormal-mode frequencies, we now
present how we obtain their \emph{excitation factors}.
The excitation factors are complex-valued constants that are characteristic
of a black-hole spacetime and partially determine the amplitude with which
different quasinormal modes are excited given an initial source of disturbance~\cite{Leaver:1986gd,Andersson:1995zk}.

In this context, we are interested in the inhomogeneous version of Eq.~\eqref{eq:general_fd_Q}
\begin{equation}
    \left[ \frac{\dd^2}{\dd x^2} + Q \right] X = s.
    \label{eq:general_fd_Q_inho}
\end{equation}
The source $s$ can represent, in the Fourier domain, either the initial data of the function $X$ in a spacelike hypersurface $t = \textrm{constant}$, or an external
source driving the perturbations $X$, for instance, a particle plunging into the black hole~\cite{Zerilli:1970wzz}.
The Cauchy problem associated to Eq.~\eqref{eq:general_fd_Q_inho} can be studied using
Green's functions~\cite{MorseFeshbach1953}. Leaver~\cite{Leaver:1986gd}, showed that the contribution from the quasinormal modes to the response in time of $X$ is given by
\begin{equation}
    X(t,r) = - {\rm Re} \,
    \sum_{n}
    \left[
        C_{n} \, \ee^{- \ii \omega_{n} (t - x)}
    \right],
\end{equation}
where the sum is over all quasinormal frequencies $\omega_n$ and
$C_n$ are the respective quasinormal excitation coefficients. The latter can be factorized as
\begin{equation}
    C_{n} = B_{n} \, I_{n},
\end{equation}
where $I_{n}$ is an integral over the source $s$ and the solution of the homogeneous
equation~\eqref{eq:eqs_rwz_fd} at the quasinormal frequency $\omega_n$,
\begin{equation}
    I_{n} = \int_{-\infty}^{\infty} \, \dd x' \,
    s(x') \, X_{n}(x') / A_{\rm out},
    \label{eq:exc_fac_source}
\end{equation}
and $B_{n}$ are the source-independent excitation factors
\begin{equation}
    B_{n} = \frac{A_{\rm out}}{2 \omega_n} \,
    \left[\frac{\dd A_{\rm in}}{\dd \omega}\right]^{-1}_{\omega = \omega_n}
    = \frac{1}{2 \omega_n} \frac{A_{\rm out}}{\alpha_n},
    \label{eq:exc_fac_def}
\end{equation}
where we approximated $A_{\rm in} \simeq \alpha_{n} (\omega - \omega_n)$ in
the vicinity of the quasinormal-mode frequency $\omega_n$.
Hence, the excitation factors are related to the ingoing- and outgoing-wave amplitudes
at spatial infinity at frequencies near $\omega_n$.

Equation \eqref{eq:exc_fac_def} is our main quantity of interest. To calculate it,
we follow Ref.~\cite{Glampedakis:2003dn} again, which proposed a phase-amplitude
based scheme to compute $B_n$; see also Ref.~\cite{Andersson:1995zk}. This means we must derive a relation between
the wave-amplitudes in Eq.~\eqref{eq:exc_fac_def} and the phase-functions $P$ and $\Pt$.
We begin by recalling that the general physical solution of Eq.~\eqref{eq:general_fd_Q}
is an ingoing wave at the event horizon and a mixture of ingoing and outgoing
waves at spatial infinity. If we were to integrate this solution first from a near-horizon location $x_{-\infty}$ up to a matching point $x_\mm$ and from a far-field location $x_\infty$ down to the same $x_\mm$,
we would express the result of these two integrations as
\begin{subequations}
\label{eq:X_LR}
\begin{align}
    X_{\rm L}(x_\mm) &= \exp(\varphi_{\LL, -}),
    \label{eq:X_L}
    \\
    X_{\rm R}(x_\mm) &=
    \mathcal{A} \exp(\varphi_{\RR, +})
    +
    \mathcal{B} \exp(\varphi_{\RR, -}),
    \label{eq:X_R}
\end{align}
\end{subequations}
where $\cA$ and $\cB$ are complex amplitudes and the various $\varphi$ are
integrals over the phases, more precisely,
\begin{equation}
    \varphi_{\LL, -}   = \int_{x_{-\infty}}^{x_\mm}  P_{\LL, -} \, \dd x',
    \quad
    \varphi_{\RR, \pm} = \int_{x_\infty}^{x_\mm} P_{\RR, \pm} \, \dd x'.
\end{equation}
In these expressions, we introduced the subscripts L (R) to indicate a
function to the left (right) of the matching point $x_\mm$, and $+$
($-$) to indicate the ingoing (outgoing) wave phase.
The condition for Eq.~\eqref{eq:X_LR} to be a solution of Eq.~\eqref{eq:general_fd_Q}
is that the logarithmic derivatives of $X_{\LL}$ and $X_{\RR}$ are continuous at $x_\mm$:
\begin{equation}
    P_{\LL, -} = \frac{P_{\RR,+} + (\cB/\cA) \, P_{\RR,-} \, \exp(\varphi_{\RR,-} - \varphi_{\RR,+}) }{1 + (\cB/\cA) \, \exp(\varphi_{\RR,-} - \varphi_{\RR,+})},
\end{equation}
which we solve for $\cB/\cA$,
\begin{equation}
    \frac{\cB}{\cA} = \frac{P_{\RR,+} - P_{\LL,-}}{P_{\LL,-} - P_{\RR,-}} \, \exp(\varphi_{\RR,-} - \varphi_{\RR,+}).
    \label{eq:BA}
\end{equation}
This is the first step of the derivation. The next step consists of finding a
relation between $\cA$ and $\cB$ with the amplitudes $A_{\rm in}$ and $A_{\rm out}$; cf.~Eq.~\eqref{eq:bcs_general}.
By doing so, as detailed in Ref.~\cite{Glampedakis:2003dn}, we can rewrite Eq.~\eqref{eq:BA} as
\begin{align}
    \frac{A_{\rm in}}{A_{\rm out}} = \frac{P_{\RR,+} - P_{\LL,-}}{P_{\LL,-} - P_{\RR,-}}
    \, \ee^{\Phi},
    \label{eq:Ain_Aout_phases}
\end{align}
where we defined
\begin{equation}
    \Phi = 2 \ii \omega x_{\infty} + \varphi_{\RR,+} - \varphi_{\RR,-}
    - 2 \ii \int_{x_{\infty}}^{\infty} \frac{Q - \omega^2}{Q^{1/2} + \omega} \, \dd x'.
    \label{eq:def_big_phase}
\end{equation}
The final step is to take a derivative of Eq.~\eqref{eq:Ain_Aout_phases} with
respect to the frequency $\omega$ and evaluate the result at the
quasinormal mode frequency $\omega_{n}$ using that
(i) $P_{\RR,+}$ is equal to $P_{\LL,-}$ at $\omega = \omega_n$ (this is nothing but the ``resonant condition'' for the quasinormal mode~\cite{Chandrasekhar:1975zza}) and
(ii) the linear approximation $A_{\rm in} \simeq \alpha_{n} (\omega - \omega_n)$.
By doing so, we obtain
\begin{align}
    \frac{\alpha_{n}}{A_{\rm out}} = \frac{\Omega_{\RR,+} - \Omega_{\LL,-}}{P_{\LL,-} - P_{\RR,-}} \,
    \ee^{\Phi_n},
    \quad \Omega_{\RR/\LL, \pm} = \frac{\dd P_{\RR/\LL,\pm}}{\dd \omega},
    \label{eq:alpha_Aout_final}
\end{align}
which is our final result~\cite{Glampedakis:2003dn}. We reiterate that all quantities in Eq.~\eqref{eq:alpha_Aout_final}
are evaluated at $x = x_\mm$ [i.e., $\rho = \rho_\mm$; see Eq.~\eqref{eq:complex_tortoise}] and $\omega = \omega_n$.
Once we have determined the value of $\alpha_n / A_{\rm out}$, we use
Eq.~\eqref{eq:exc_fac_def} to calculate the excitation factor $B_n$
of the quasinormal mode frequency $\omega_n$.

How do we calculate the various terms entering Eqs.~\eqref{eq:alpha_Aout_final}
and~\eqref{eq:def_big_phase}? From these equations we identify two terms that
are independent on the phase functions, namely
\begin{equation}
    2 \ii \omega_{n} x_\infty,
    \quad \textrm{and} \quad
    {\mathcal I} = - 2 \ii \int_{x_{\infty}}^{\infty} \frac{Q - \omega_n^2}{Q^{1/2} + \omega_n} \, \dd x'.
\end{equation}
The former is a constant, while the latter can be integrated analytically by first expanding the integrand
in powers of $1/x \approx 1/r$  (since $|x/\rh| \gg 1$) and then
integrating term by term. The integral is convergent in our case,
for which $Q \simeq \omega^2$ and $\dd Q / \dd x \simeq 0$ as $x \to \infty$.
We find:
%
%
\begin{equation}
    {\mathcal I} = - \frac{\ii}{2 \omega}
    \left[
    u_{\infty} \left. \frac{\dd^2Q}{\dd u^2} \right\vert_{u=0}
    +
    \frac{u^2_{\infty}}{6} \left. \frac{\dd^3 Q}{\dd u^3} \right\vert_{u=0}
    +
    \dots
    \right],
\end{equation}
where $u = 1 / x$.

For the remaining terms, it is convenient to separate our discussion into quantities that are determined between $x \in [x_{-\infty}, x_\mm]$ and $x \in (x_\mm, x_{\infty}]$.
Because we must evaluate the phase functions at a
quasinormal frequency, it is useful to use
the analytical continuation and the Pr\"ufer
transformation introduced in Sec.~\ref{sec:phase_amp}.
By doing so, we can write down two systems of
first-order differential equations.
Specifically, for  $x \in [x_{-\infty}, x_\mm]$, we need to integrate three equations
\begin{subequations}
\label{eq:exc_fac_region_horizon}
\begin{align}
    &\frac{\dd P_{\LL,-}}{\dd \rho} + \ee^{\ii \beta} (P^2_{\LL,-} + Q) = 0, \\
    &\frac{\dd \Omega_{\LL,-}}{\dd \rho} + \ee^{\ii \beta}
    \left[ 2 P_{\LL,-} \, \Omega_{\LL,-} + \frac{\dd Q}{\dd \omega} \right] = 0,
     \\
    &\frac{\dd r}{\dd \rho} - \ee^{\ii \beta} \frac{\dd r}{\dd x} = 0,
\end{align}
\end{subequations}
with initial conditions at $\rho = \rho_{\min}$ given by
\begin{equation}
    P_{\LL,-} = \ii Q^{1/2},
    \quad \textrm{and} \quad
    \Omega_{\LL,-} = - \frac{1}{2 P_{\LL,-}} \frac{\dd Q}{\dd \omega}.
\end{equation}
For $x \in (x_\mm, x_{\infty}]$, we need to integrate six equations
\begin{subequations} \label{eq:exc_fac_region_inf}
\begin{align}
    &\frac{\dd \Pt_{\RR,\pm}}{\dd \rho} + \ee^{\ii \beta}
    \left(\omega - \frac{Q}{\omega} \right)
    \sin^2 (\Pt_{\RR,\pm} + |\omega| \rho + \omega x_{\mm} ) = 0, \\
    &\frac{\dd \varphi_{\RR,\pm}}{\dd \rho}  -
    \ee^{\ii \beta} \omega \cot(\omega x + \Pt_{\RR,\pm}) = 0,  \\
    &\frac{\dd \Omega_{\RR,+}}{\dd \rho} + \ee^{\ii \beta}
    \left[ 2 \, \Omega_{\RR,+} \, \omega \cot(\omega x + \Pt_{\RR,+})
    + \frac{\dd Q}{\dd \omega} \right] = 0,  \\
    &\frac{\dd r}{\dd \rho} - \ee^{\ii \beta} \frac{\dd r}{\dd x} = 0.
\end{align}
\end{subequations}
The initial conditions for $\varphi_{\RR,\pm}$ and $\Omega_{\RR,+}$ at $\rho = \rho_{\rm max}$ are
\begin{align}
\varphi_{\RR,\pm} = 0,
\quad \textrm{and} \quad
\Omega_{\RR,+} = - \frac{1}{2 P_{\RR,+}} \frac{\dd Q}{\dd \omega},
\end{align}
and to determine the initial condition for $\Pt_{\RR,\pm}$, we first use the
WKB formula $P_{\RR,\pm} = \pm \ii Q^{1/2}$ and then substitute the result in Eq.~\eqref{eq:prufer}.
In our case $\dd Q / \dd \omega = 2 \omega c^{-2}_{\rm s}$.

Equations~\eqref{eq:exc_fac_region_horizon} and~\eqref{eq:exc_fac_region_inf}
constitute the two systems of differential equations we need to integrate to
compute the quasinormal-mode excitation factors.
The integration procedure can be summarized as follows:
\begin{enumerate}
    \item Choose the values of $\varepsilon$, $\ell$, and \emph{quasinormal mode frequency}
    $\omega_n$, and with the latter compute $\beta = - \arg \omega_n$.
    \item Determine the initial conditions for $r$ as $\rho \to \pm \infty$.
    We integrate Eq.~\eqref{eq:dr_drho} from $\rho_\mm = 0$
    backward (forward) to, say, $\rho_{\rm min} = - 40M$
    ($\rho_{\rm max} = 2 \times 10^{4}M$). This fixes
    $r_{\rm min} = r(\rho_{\rm min})$ and $r_{\rm max} = r(\rho_{\rm max})$.
    \item Determine the other initial conditions for the dependent variables in
    the two integration domains, that is, at $\rho_{\rm min}$ and $\rho_{\rm max}$.
    \item Integrate the system of equations~\eqref{eq:exc_fac_region_horizon} and~\eqref{eq:exc_fac_region_inf} from $\rho_{\rm min}$ to $\rho_\mm$
    and from $\rho_{\rm max}$ to $\rho_\mm$, respectively.
    \item Calculate $P_{\RR,\pm}(\rho_{\mm})$ from $\Pt_{\RR,\pm}(\rho_{\mm})$ using Eq.~\eqref{eq:phase_from_prufer}.
    \item Compute $\alpha_n / A_{\rm out}$ using Eqs.~\eqref{eq:def_big_phase} and \eqref{eq:alpha_Aout_final}, and, finally, the excitation factor $B_{n}$ using Eq.~\eqref{eq:exc_fac_def}.
\end{enumerate}
We implemented the foregoing steps in C++, adopting the same integration
library as in our calculation of the quasinormal mode frequencies.
We will present our results for the quasinormal-mode excitation factors in Sec.~\ref{sec:exc}.

\section{Numerical results}
\label{sec:results}

\subsection{The quasinormal mode spectrum}
\label{sec:qnm}

\subsubsection{Comparison with the literature}
\label{sec:results:qnmcomparison}

Our calculation of the quasinormal-mode frequencies was validated in two ways.
First, in the limit of general relativity, we compared our results against the well-known
values for a Schwarzschild black hole, finding excellent agreement. Computation wise,
we found that it was necessary to shift the matching point closer to the event horizon
to accurately calculate the overtone frequencies. Above a certain overtone number, typically $n \gtrsim 4$, it becomes increasingly challenging to locate the quasinormal mode in the root-finding process.
The reason is that the simple integration path~\eqref{eq:complex_tortoise} fails to approximate the more complex integration path necessary for determining high overtone quasinormal frequencies; see~Refs.~\cite{Andersson:1992scr,Andersson:1993CQGra}. For this reason,
we quote results up to $n=3$.

\begin{figure}[t]
\includegraphics{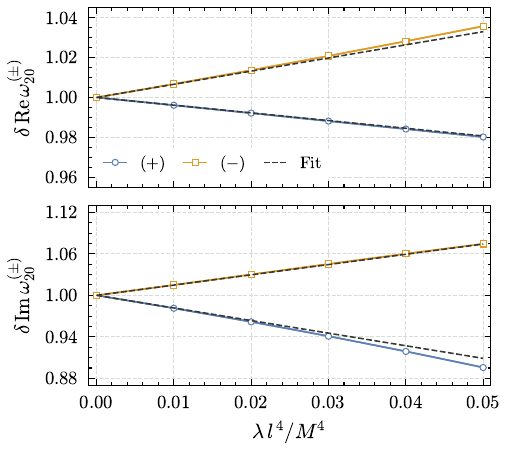}
\caption{The fundamental polar and axial quasinormal mode frequencies,
normalized with respect to their Schwarzschild values, as functions of
$\varepsilon = \lambda \, l^4 / M^4$. The top panel shows the real part of
frequencies, while the bottom panel their imaginary part. The markers distinguish curves corresponding to modes of polar ($+$) and axial ($-$) parities. The dashed lines are the linear fits by Cano~et~al.~\cite{Cano:2023jbk}. Both calculations agree remarkably well despite using two different forms of the perturbations equations to which different numerical techniques were applied to compute the quasinormal mode frequencies.}
\label{fig:comparison_against_cano_etal}
\end{figure}

\begin{figure*}[ht]
\includegraphics{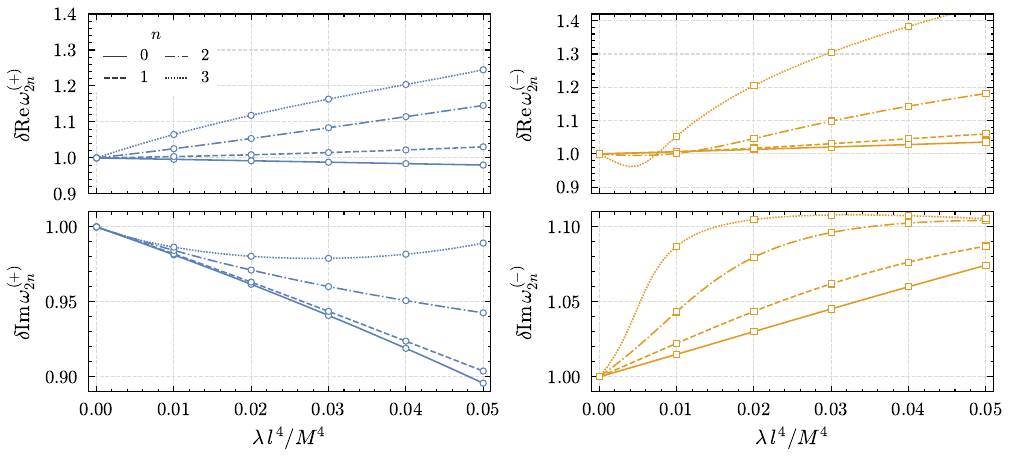}
\caption{The quadrupolar polar- and axial-parity quasinormal frequencies,
normalized with respect to their Schwarzschild values, as functions of
$\varepsilon = \lambda \, l^4 / M^4$. The left and right columns correspond
to quasinormal modes of polar and axial parities, respectively, whereas the top and bottom panels show the real and imaginary parts.
The line styles indicate different overtone numbers $n$.
We see that the deviations from the general-relativistic values can become nonmonotonic
as we increase the value of $\varepsilon$. We also find that the curves become nonlinear
for smaller values of $\varepsilon$ the higher the overtone number $n$.}
\label{fig:rel_change_l2}
\end{figure*}

As an example, for the fundamental and third overtone quadrupole
quasinormal frequencies we obtain
\begin{subequations}
\begin{align}
    M\omega^{(+)}_{20} &= 0.37367169-0.088962318 \ii,
    \\
    M\omega^{(-)}_{20} &= 0.37367169-0.088962321 \ii,
\end{align}
\end{subequations}
and
\begin{subequations}
\begin{align}
    M\omega^{(+)}_{23} &= 0.25150495-0.70514814 \ii,
    \\
    M\omega^{(-)}_{23} &= 0.25150496-0.70514818 \ii,
\end{align}
\end{subequations}
respectively. By isospectrality, we expect $\omega^{(+)}_{\ell n} = \omega^{(-)}_{\ell n}$, and, indeed, the phase-amplitude method yields quasinormal frequencies that differ from one another by ${\cal O}(10^{-6})$ or better.
In addition, we compared our results against those of the \texttt{qnm} package~\cite{Stein:2019mop} that uses the continued-fraction method of Refs.~\cite{Leaver:1985ax,Cook:2014cta}, and gives:
\begin{subequations}
\begin{align}
    M\omega^{(\pm)}_{20} &= 0.37367168-0.088962316 \ii,
    \\
    M\omega^{(\pm)}_{23} &= 0.25150496-0.70514820  \ii.
\end{align}
\end{subequations}
We find relative errors of ${\cal O}(10^{-6})$ that are largest for
the highest overtone, $n=3$, for the reason explained earlier.
After having computed the general-relativistic
quasinormal frequencies to good accuracy, we varied the EFT parameter $\varepsilon$
in constant steps of $\Delta \varepsilon = 0.001$ and scanned the domain $\varepsilon \in [0, 0.05]$.
In Figs.~\ref{fig:comparison_against_cano_etal} and~\ref{fig:rel_change_l2}, we
use markers to indicate values of $\varepsilon$ in increments of $10 \times \Delta \varepsilon = 0.01$ to ``guide the eye.''

In Fig.~\ref{fig:comparison_against_cano_etal} we show the
real and imaginary parts of the fundamental quadrupolar quasinormal frequency as a function of $\varepsilon$ and normalized with respect to its value in general relativity, i.e.,
\begin{equation}
    \delta \real \omega^{(\pm)}_{\ell n}(\varepsilon) =
    \frac{\real \omega^{(\pm)}_{\ell n}(\varepsilon)}{\real \omega^{(\pm)}_{\ell n}(0)},
    \quad
    \delta \imag \omega^{(\pm)}_{\ell n}(\varepsilon) =
    \frac{\imag \omega^{(\pm)}_{\ell n}(\varepsilon)}{\imag \omega^{(\pm)}_{\ell n}(0)}. \nn
\end{equation}
The top and bottom panels show our results for these two quantities, respectively.
The solid curves are the results
of our phase-amplitude calculation. We use the markers to distinguish the curves corresponding to quasinormal modes of polar ($+$) and axial ($-$)
parities, which are no longer the same.
We also show, with the dashed curves, the fits obtained in Ref.~\cite{Cano:2023jbk}; cf.~Table~III therein. These fits are a
linear approximation to the behavior of $\omega^{(\pm)}_{20}$ with respect to $\varepsilon$, which does become nonlinear as $\varepsilon$ growth; cf.~Ref.~\cite{Cano:2021myl}, Fig.~1. The same behavior can be seen here.
In addition, our results are in excellent agreement with the
linear fits of Ref.~\cite{Cano:2023jbk} for $\varepsilon \lesssim 0.03$
and across the whole $\varepsilon$ range for $\delta \real \omega^{(+)}_{20}$ and $\delta \imag \omega^{(-)}_{20}$.
The agreement is quite remarkable considering that we do not integrate the same set of perturbations equations and that we use different numerical techniques to compute the quasinormal modes; recall Sec.~\ref{sec:bhpt}.
We find a similar level of agreement for the fundamental $\ell = 3$ quasinormal mode frequency.
As a consequence, our numerical results are also in agreement
with those of de Rham et al.~\cite{deRham:2020ejn}.

We also briefly studied the case where $\varepsilon$ is negative. For
sufficiently small values of $|\varepsilon|$, we would expect that the
deviations from the quasinormal frequencies in general relativity to be equal
in magnitude, but with an opposite sign relative to the case where
$\varepsilon$ is positive. We found that this was indeed the case for
$\omega_{20}^{(\pm)}$.

\subsubsection{Overtones and the limits of the effective field theory}
\label{sec:results:overtones}

\begin{figure*}[t]
\includegraphics{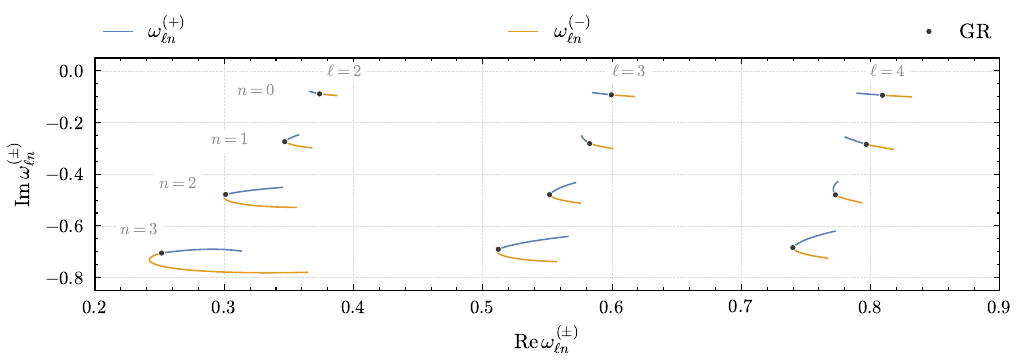}
\caption{The spectrum of quasinormal modes $\omega_{\ell n}^{(\pm)}$ in the complex plane
in the range $\varepsilon \in [0,\, 0.05]$.
The circles mark the location of the general-relativity (``GR'') quasinormal mode frequencies,
that are coincident for axial and polar modes. A nonzero value of the EFT parameter $\varepsilon$
breaks this symmetry, and axial- and polar-parity frequencies flow along the
blue and orange colored lines, respectively, as we increase $\varepsilon$.
The larger the overtone number $n$, the farther away the quasinormal frequencies, at fixed multipole number $\ell$, with respect to the fundamental mode, $n=0$.
}
\label{fig:cp}
\end{figure*}

With confidence built on the applicability of the phase-amplitude approach to
our problem and on our numerical code, we now investigate, for the first time,
the dependence of the overtones as a function of $\varepsilon$.

In Fig.~\ref{fig:rel_change_l2} we show $\omega^{(\pm)}_{2n}$ as a function of $\varepsilon$. Panels on the left and right columns show the polar ($+$) and axial ($-$) quasinormal frequencies, respectively. We show their real and imaginary parts (normalized with respect to their values in general relativity) in the top and bottom rows, respectively. Different line styles represent different overtone numbers $n$ as indicated in the legend in the top-left panel.
We observe that as the overtone number $n$ increases, the smaller the range of $\varepsilon$ at which the scaling is linear. Moreover the curves do not necessarily behave monotonically with respect to $\varepsilon$. This can best be seen for the $n=3$ axial-parity quasinormal-mode frequency.
These conclusions are shared among the overtones associated to
$\ell = 2$, 3 and 4. The emerging picture has two facets:
(i) \emph{overtones are more sensitive to the EFT corrections than the fundamental mode},
and
(ii) \emph{the maximum value of $\varepsilon$ above which the linear approximation breaks down depends on $\ell$ and $n$}.

In order to discuss the behavior of the quasinormal-mode frequencies, we first describe the regime of applicability of the EFT of gravity.
As already mentioned, the EFT corrections can be computed in powers of $\varepsilon$, with the EFT breaking down once $\varepsilon \simeq \varepsilon_{\rm th}$ where $\varepsilon_{\rm th}$ is a threshold value of order 1. This is the statement that the black-hole curvature radius has to be larger than the scale $l$.
Moreover, one also needs the frequency of perturbations not too large.
At fixed $\ell$, increasing $n$ corresponds in increasing the proper frequency $f_{\ell n}$ of the quasinormal modes. The quantity $f_{\ell n}$ can be identified with the real part of $\omega_{\ell n} / (2 \pi)$ only if ${\rm Re \,} \omega_{\ell n} \gg {\rm Im \,} \omega_{\ell n}$. Typically this is not the case for the Schwarzschild quasinormal modes and, following Ref.~\cite{Maggiore:2007nq}, one has to make the following identification in order to have a monotonically increasing spectrum of proper frequencies
\begin{equation}
f_{\ell n} = (2\pi)^{-1} \,[\, \real \omega^2_{\ell n} + \imag \omega^2_{\ell n} \,]^{1/2}.
\label{eq:proper_frequency}
\end{equation}
Then, at fixed $\varepsilon$, an overtone with proper frequency $f_{\ell n}$ can be described within the EFT provided that \cite{Chen:2021bvg}
\begin{equation}\label{eq:def_eps_f}
    \varepsilon_{f} = \lambda \, (l f_{\ell n})^4  = \varepsilon \, ( M f_{\ell n})^4 \ll \varepsilon^{-1}.
\end{equation}
The condition above is the statement that the covariant contraction $k^{\mu}
\kappa_{\mu} \ll l^{-2}$, where $k^{\mu}$ and $\kappa^{\mu}$ are respectively
the typical four-momenta of the gravitational perturbations and of the black
hole background. The latter can be defined as the normalized covariant
derivative of the Kretschmann scalar~\cite{Chen:2021bvg}.

After these considerations, let us discuss the behavior we find for the quasinormal modes.
Our results show a growing impact of the EFT corrections on $\omega_{\ell n}$
as $n$ increases at a fixed multipole $\ell$; see the ``bird's-eye view'' shown
in Fig.~\ref{fig:cp}.
Reasoning in terms of Eq.~\eqref{eq:proper_frequency} offers a
qualitative explanation for this behavior. Overtones correspond to high-frequency waves and consequently probe deeper the effective potential which is fixed by $\varepsilon$ and $\ell$. As a consequence, overtones are more sensitive
to changes to the effective potential that occur near the black-hole horizon,
which is the case in our problem --- recall Fig.~\ref{fig:potentials}.
Complementary, let us also remark that this behavior is a general feature of perturbation theory when the correction to the potential comes from regions far from the potential peak, or the potential minimum in the bound-problem case. An example in quantum mechanics is given by a harmonic oscillator potential $V(x) \propto x^2$ in the presence of a small anharmonicity (going as, e.g.~$\delta V(x) \propto x^4$).
At first order in perturbation theory, the corrections to the $n$th eigenenergy grow as $n$ to some given power (see for example Ref.~\cite{CohenQM}).
In our case, the EFT corrections grow toward the horizon while the peak of the unperturbed potential is approximately at the light ring and we therefore expect a similar behavior.

The nonlinear behavior above a certain $\varepsilon_{\rm max}$
of the quasinormal frequencies sets an upper limit on the
linear description of our problem, and indicates that higher powers of
$\varepsilon$ are necessary to describe the regime for
$\varepsilon \gtrsim \varepsilon_{\rm max}$.
This means that we either need to go to second order in perturbation theory (our perturbations equations are linear in the metric perturbations and in $\varepsilon$) or that we need to include higher-order operators in our starting action~\eqref{eq:action_schematic}.
At second order in perturbation theory the second-order quasinormal frequencies are a sum
of first-order quasinormal frequencies~\cite{Ioka:2007ak,Nakano:2007cj}; see, e.g., Refs.~\cite{Pazos:2010xf,London:2014cma,Loutrel:2020wbw,Ripley:2020xby,Cheung:2022rbm,Mitman:2022qdl,Bucciotti:2023ets} for further details. As a consequence, we expect the second-order
quasinormal frequencies to also scale with $\varepsilon$ in the EFT.
Assuming that all the dimensionless factors entering in the higher-dimension
operators of the EFT are numbers of order one, which is technically
natural~\cite{Endlich:2017tqa}, then the effects from dimension-eight operators
would be of order $\varepsilon^{3/2}$.
Hence, they would be the dominant contribution to the quasinormal
frequencies in the nonlinear regime of $\varepsilon$.

\begin{figure}[ht]
\includegraphics{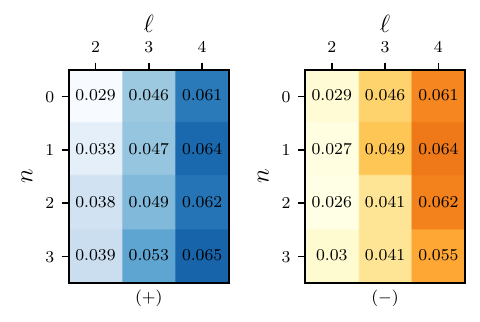}
\caption{Values of $\varepsilon_f^{1/4}$ at which a given quasinormal frequency deviates by more than $20\%$ from its linear fit in $\varepsilon$. Polar $(+)$ and axial $(-)$ quasinormal modes are shown on the left and right panels, respectively. Notice that the values of $\varepsilon_f^{1/4}$ in the tables vary only by a factor of two between different modes.}
\label{fig:epsilon_f}
\end{figure}

From these results and considerations, we find that the onset of the nonlinear behavior in the quasinormal-mode frequencies is best characterized in terms of the parameter $\varepsilon_f$, rather than $\varepsilon$.
In Fig.~\ref{fig:epsilon_f} we show, for each $\ell$ and $n$, the value of $\varepsilon_f^{1/4}$ at which nonlinear corrections start appearing.
More specifically, we first evaluate the value of $\varepsilon$ at which $\omega^{(\pm)}_{\ell n}$ deviates from its linear fit by more than $20\%$.
This gives us values of $\varepsilon_{\rm max}$ and $\omega_{\ell n, {\rm max}}^{(\pm)} = \omega^{(\pm)}_{\ell n}(\varepsilon_{\rm max})$ associated to this mode \emph{and} threshold value.
For instance, for $\omega^{(\pm)}_{2 0}$ we find $\varepsilon_{\rm max} > 0.05$, hence all values of $\varepsilon$ considered by us are interpreted
to be within the linearized regime according to this criteria. In this case, we take for $\omega_{\ell n, {\rm max}}^{(\pm)}$ the value of $\omega_{\ell n}^{(\pm)}$ at $\varepsilon = 0.05$. However, in general,
this is not what happens for other values of $\ell$ and $n$.
Then, with the values of $\varepsilon_{\rm max}$ and $\omega_{\ell n, {\rm max}}^{(\pm)}$ in hand,
we can compute the respective value of
$\varepsilon_f$ using Eq.~\eqref{eq:def_eps_f}.
The values obtained are below one, meaning that the onset of nonlinearities appears below the breaking of the EFT, as one would expect.

\begin{figure}[t]
\includegraphics[scale=0.5]{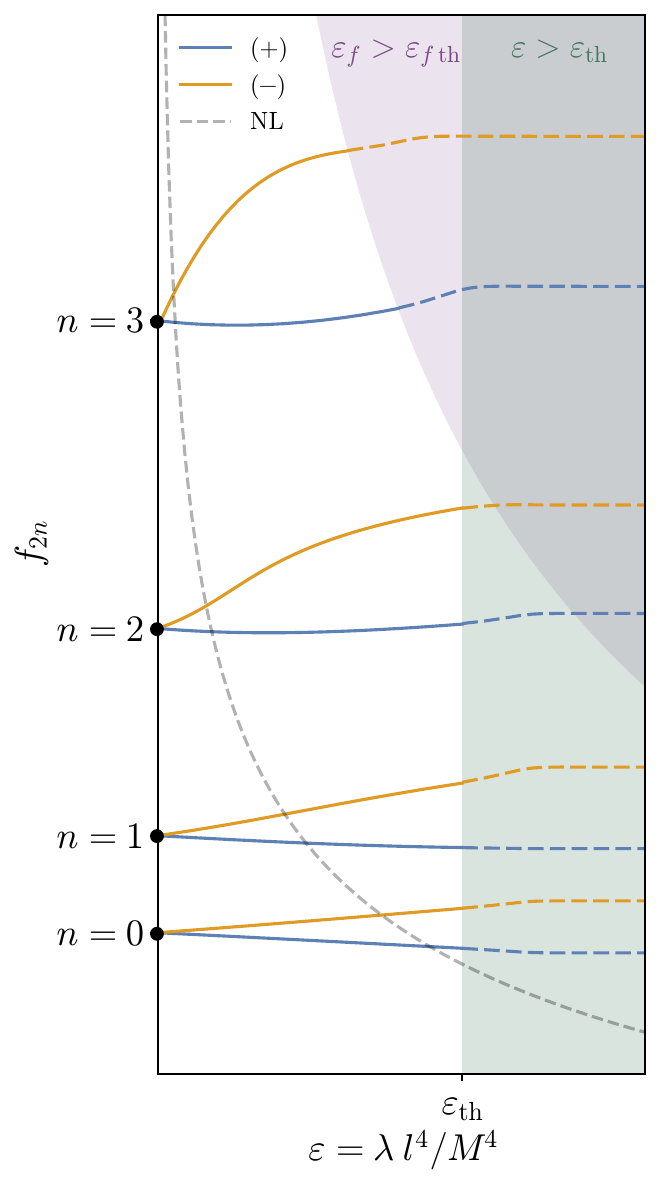}
\caption{
Schematic representation of the quadrupolar quasinormal proper frequencies $f_{2 n}$
for even (blue) and odd (orange) parities, as functions of $\varepsilon$ and the regions of validity of the EFT calculation.
For small $\varepsilon$, deviations from the GR values are approximately linear. The onset of nonlinearities in the corrections is represented by the gray dashed line (NL): above this line one needs to include higher-order contributions in $\varepsilon$.
When $\varepsilon \gtrsim \varepsilon_{\rm th} \sim \mathcal{O}(1)$ (green-shaded area) or $\varepsilon_{f} \gtrsim  \varepsilon_{f \, {\rm th}} \sim \varepsilon^{-1}$ (purple-shaded area) the EFT description breaks down and one has to resort to its UV completion to make predictions; see the discussion around Eq.~\eqref{eq:def_eps_f}. Under the assumption of a soft UV completion, the corrections to the quasinormal frequencies are expected to saturate, as illustrated with the dashed lines.
}
\label{fig:spectrum_soft_uv}
\end{figure}

As mentioned above, for values of both $\varepsilon$ and $\varepsilon_f$ close to unity the EFT description inevitably breaks down: it is not enough to include higher-order operators and any prediction can only be made with an ultraviolet (UV) completion of the EFT. As explained in Ref.~\cite{Endlich:2017tqa}, assuming a \emph{soft UV completion}, the corrections to the quasinormal-mode frequencies (and other observable quantities) are expected to saturate after the breaking of the EFT.
In Fig.~\ref{fig:spectrum_soft_uv} we give a schematic representation of the behavior of $f_{\ell n}$ in this scenario; see also Fig.~1 of Ref.~\cite{Sennett:2019bpc} and related discussion therein.

At this point it is clear that the condition in Eq.~\eqref{eq:def_eps_f} is necessarily violated for sufficiently large $n$, given a value of $\varepsilon$. In other words the EFT \emph{cannot describe all overtones}. Using the asymptotic behavior of the Schwarzschild quasinormal frequencies in general relativity we can get a good estimate for this maximum value of $n$, that we call $n_{\rm max}$. The spectrum at large $n$ is independent on $\ell$ and is given by
\begin{equation}
    M \omega_{\ell n} \simeq (8 \pi)^{-1} \log 3 - \ii \, ( n + 1/2) / 4, \quad \textrm{as} \quad n \to \infty.
\end{equation}
See Refs.~\cite{Nollert:1993zz,Andersson:1993CQGra} for numerical
studies in this limit and Refs.~\cite{Motl:2002hd,Motl:2003cd,MaassenvandenBrink:2003as,Andersson:2003fh}
for the posterior analytical derivations of this result.
Using this expression, Eqs.~\eqref{eq:proper_frequency} and \eqref{eq:def_eps_f}, and imposing the latter as an equality, we readily obtain
\begin{equation}
n_{\rm max} \simeq (8 \pi )  / \sqrt{\varepsilon} - 1/2 + \mathcal{O}
(\sqrt{\varepsilon}).
\end{equation}
The value we obtained is quite large even for $\varepsilon$ close to unity, $n_{\rm max} \sim \mathcal{O}(25)$.
Corrections not captured by the EFT are bound to appear below, or at most at, $n_{\rm max}$.

\subsection{The quasinormal-mode excitation factors}
\label{sec:exc}

\subsubsection{Code validation and comparison with the literature in the limit of general relativity}

To our knowledge this is the first time that quasinormal-mode excitation factors have been computed
for black-hole solutions in a theory that is not general relativity.
To validate our numerical results in the limit of general relativity we performed two tests.

First, in general relativity, the excitation factors of polar and axial
quasinormal modes are related as
\begin{equation}
    B^{\ps}_{\ell n} =
    \frac{2\lambda_\ell(\lambda_\ell + 1) + 6 \ii M \omega_{\ell n}}{2\lambda_\ell(\lambda_\ell + 1) - 6 \ii M \omega_{\ell n}} \,
    B^{\mn}_{\ell n},
    \label{eq:leaver_relation}
\end{equation}
as shown by Leaver~\cite{Leaver:1986gd}.
This relation follows from two identities relating the transmission and reflection
coefficients of the Zerilli and Regge-Wheeler functions found by Chandrasekhar
and Detweiler~\cite{Chandrasekhar:1975zza}.
We note that despite sharing the same quasinormal mode spectra,
the Zerilli and Regge-Wheeler modes have different excitation factors.
As a consistency check of our code, we verified that
our numerical calculations of $B^{(\pm)}_{\ell n}$ satisfy Eq.~\eqref{eq:leaver_relation} with a mean error of approximately ${\cal O}(10^{-5})$
across the $\ell n$-parameter space we studied. This error decreases by one order of magnitude if we exclude the $n=3$ overtones.

As a second test, we compared our values of $B^{(\pm)}_{\ell n}$ with those obtained
by Zhang, Berti, and Cardoso~\cite{Zhang:2013ksa} (see Table II therein), who employed
the formalism of Mano, Suzuki, and Takasugi~\cite{Mano:1996mf,Mano:1996vt,Fujita:2004rb}. This scheme is based
on a matched asymptotic expansion between a Coulomb-series expansion near spatial infinity and a series expansion
involving hypergeometric functions near the event horizon. Reassuringly, we find excellent
agreement between our phase-amplitude-based calculation and those of Ref.~\cite{Zhang:2013ksa}.
For example, for the excitation factors of the fundamental quadrupole quasinormal frequencies
we have
\begin{subequations}
\begin{align}
    B^{(+)}_{20} &= 0.120928+0.0706657 \ii,
    \\
    B^{(+)}_{20} &= 0.120923 + 0.0706696 \ii,
\end{align}
\end{subequations}
for the polar-parity mode and
\begin{subequations}
\begin{align}
    B^{(-)}_{20} &= 0.126902 + 0.0203152 \ii,
    \\
    B^{(-)}_{20} &= 0.126902 + 0.0203152 \ii,
\end{align}
\end{subequations}
for the axial-parity mode. In each of the two foregoing equations the second line is taken from Ref.~\cite{Zhang:2013ksa}, Table II.

\begin{table*}
\begin{tabular}{c c c c c}
\arrayrulecolor{Gray}
\hline
\hline
$B^{(+)}_{\ell n}$ & $\varepsilon \, (\times 10^{-2})$ & $\ell = 2$ & $\ell =3$ & $\ell = 4$ \\ \hline
        & 0 & $0.120928+0.0706657 \ii$ & $-0.0889688-0.0611773 \ii$ & $0.0621245+0.069099 \ii$ \\
$n = 0$ & 1 & $0.117708+0.0672479 \ii$ & $-0.087801-0.0568295 \ii$ & $0.0628948+0.0644908 \ii$ \\
        & 2 & $0.11423+0.0639308 \ii$ & $-0.0863858-0.0524486 \ii$ & $0.0634276+0.0597634 \ii$ \\
\hline
        & 0 & $0.158645-0.253326 \ii$ & $-0.191931+0.264798 \ii$ & $0.279718-0.24183 \ii$ \\
$n = 1$ & 1 & $0.164761-0.235166 \ii$ & $-0.183766+0.246502 \ii$ & $0.259492-0.229203 \ii$ \\
        & 2 & $0.171749-0.217066 \ii$ & $-0.176498+0.227387 \ii$ & $0.239777-0.215239 \ii$ \\
\hline
        & 0 & $-0.298938-0.0711347 \ii$ & $0.43677+0.20459 \ii$ & $-0.543165-0.478076 \ii$ \\
$n = 2$ & 1 & $-0.298122-0.115973 \ii$ & $0.394863+0.230675 \ii$ & $-0.478737-0.464904 \ii$ \\
        & 2 & $-0.297456-0.16154 \ii$ & $0.353727+0.258994 \ii$ & $-0.413138-0.455013 \ii$ \\
\hline
        & 0 & $0.11382+0.204126 \ii$ & $-0.000943158-0.476399 \ii$ & $-0.374548+0.859556 \ii$ \\
$n = 3$ & 1 & $0.0959369+0.280675 \ii$ & $0.0877723-0.47838 \ii$ & $-0.456808+0.756741 \ii$ \\
        & 2 & $0.076331+0.367397 \ii$ & $0.177021-0.481374 \ii$ & $-0.542232+0.6564 \ii$ \\
\hline \hline
$B^{(-)}_{\ell n}$ & $\varepsilon \, (\times 10^{-2})$ & $\ell = 2$ & $\ell =3$ & $\ell = 4$ \\ \hline
        & 0 & $0.126902+0.0203152 \ii$ & $-0.0938897-0.0491928 \ii$ & $0.0653479+0.0652391 \ii$ \\
$n = 0$ & 1 & $0.130744+0.0222580 \ii$ & $-0.095795-0.0519542 \ii$ & $0.0659238+0.0680835 \ii$ \\
        & 2 & $0.134644+0.0251759 \ii$ & $-0.097599-0.0548384 \ii$ & $0.0664031+0.0709206 \ii$ \\
\hline
        & 0 & $0.047682{7}-0.223755 \ii$ & $-0.15113{5}+0.269749 \ii$ & $0.26148{8}-0.251524 \ii$ \\
$n = 1$ & 1 & $0.0{389361}-0.24{2548} \ii$ & $-0.15{4331}+0.28{9785} \ii$ & $0.27{4625}-0.26{8574} \ii$ \\
        & 2 & $0.03{30242}-0.26{2936} \ii$ & $-0.1{59411}+0.30{9949} \ii$ & $0.2{89045}-0.28{4968} \ii$ \\
\hline
        & 0 & $-0.190283+0.0157516 \ii$ & $0.415029+0.141039 \ii$ & $-0.54921{6}-0.435328 \ii$ \\
$n = 2$ & 1 & $-0.{199520}+0.0{607662} \ii$ & $0.45{8619}+0.1{08671} \ii$ & $-0.61{9525}-0.4{34249} \ii$ \\
        & 2 & $-0.2{19544}+0.09{42148} \ii$ & $0.50{6304}+0.0{827037} \ii$ & $-0.6{9088}-0.4{39270} \ii$ \\
\hline
        & 0 & $0.080858{6}+0.079601{9} \ii$ & $-0.04340{27}-0.41274{8} \ii$ & $-0.31692{2}+0.8379{11} \ii$ \\
$n = 3$ & 1 & $0.1{30958}+0.03{53271} \ii$ & $-0.13{8089}-0.4{18874} \ii$ & $-0.2{17410}+0.94{4489} \ii$ \\
        & 2 & $0.17{3338}+0.02{40910} \ii$ & $-0.2{20413}-0.4{44001} \ii$ & $-0.1{31563}+1.0{5946} \ii$ \\
\hline \hline
\end{tabular}
\caption{Quasinormal-mode excitation factors of the Zerilli and Regge-Wheeler functions for a selection
of multipoles $\ell$ overtones $n$ numbers and $\varepsilon = \lambda \, l^4/M^4 = 0$, 0.01 and 0.02.
Our results in the limit of general relativity, $\varepsilon = 0$, are in excellent agreement with the calculations of Ref.~\cite{Zhang:2013ksa}. }
\label{tab:exc_fac}
\end{table*}

\subsubsection{Effective-field-theory corrections to the excitation coefficients}

\begin{figure}[t]
\includegraphics{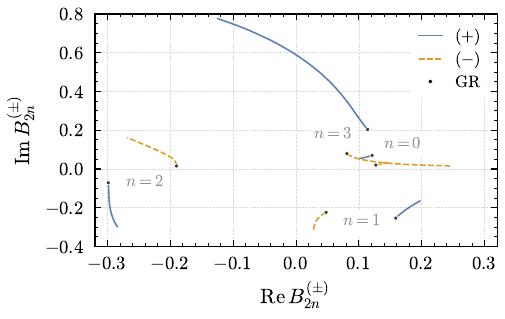}
\caption{Quadrupolar quasinormal mode excitation factors $B^{(\pm)}_{2n}$ in the range $\varepsilon \in [0,0.05]$. The circles
mark the limit of general relativity. Solid and dashed lines correspond to
the excitation factors of the quasinormal modes of polar and axial parities, respectively. The labels indicate the pairs of curves that are associated to each overtone number $n$. We see that the excitation factors move farther away
from their general-relativistic values the larger the overtone number. The same behavior occurs for the higher multipoles $\ell$ we studied.}
\label{fig:cp_exc_fac}
\end{figure}

As we observed for the quasinormal-modes frequencies, we also
find that the associated excitation factors vary more
with respect to their values in general relativity, as functions of
$\varepsilon$, for the overtones.
As an example, we show in Fig.~\ref{fig:cp_exc_fac} the trajectories in the complex plane of the quadrupolar quasinormal mode excitation factors as we increase $\varepsilon$ from zero (circles) to 0.05.
Solid and dashed lines correspond to the excitation factors of the quasinormal
modes of polar and axial parities, respectively. Pairs of curves belonging
to the same overtone number $n$ are indicated by the labels.
We found the same qualitative behavior for higher multipoles $\ell$.
We present some sample values of the excitation factors in Table~\ref{tab:exc_fac}.

\begin{figure}[b]
\includegraphics{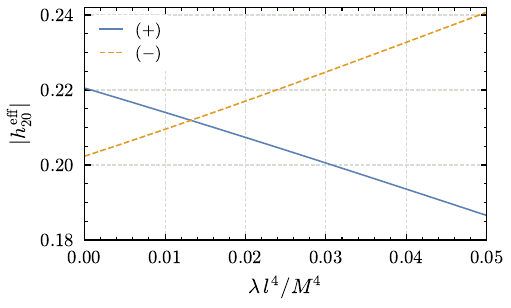}
\caption{The effective quasinormal mode amplitude for the fundamental quadrupolar quasinormal modes of axial $(-)$ and polar $(+)$ parities as a functions of $\varepsilon = \lambda \, l^4/M^4$. For $\varepsilon \lesssim 0.012$, the amplitude is largest for the for polar quasinormal mode. A crossover occurs around $\varepsilon\approx 0.012$, above which the amplitude of the axial quasinormal mode becomes larger.}
\label{fig:heff}
\end{figure}

\subsubsection{The effective quasinormal mode amplitude}

As an application of the calculations presented in this section,
we perform a preliminary analysis of the gravitational wave amplitude
associated with the polar and axial quasinormal modes. In practice this is done
by introducing an ``effective'' amplitude related to the magnitude of the Green's function
used in the solution of the radial perturbation equations for a given set of initial data; see, e.g.,~Ref.~\cite{Glampedakis:2001js}.
The relevant part of the Green's function is the
one describing the quasinormal mode ringdown signal and is given by $h = A_{\rm out} (\omega_n)/\alpha_n $
for each individual mode $\omega_n$. The effective amplitude is then $h_{\rm eff} = \sqrt{N} \, h$, where
$N = \sqrt{\real \omega_n / \imag\omega_n}$ is the number of cycles in the ringdown signal; this obviously
assumes a weakly damped mode.
Therefore, the effective signal amplitude of each polar and axial quasinormal mode
$\omega_{\ell n}$ is given by
\begin{equation}
    h^{\,(\pm),\,{\rm eff}}_{\ell n} =
    2 \, \left[ \frac{\real \omega^{\,(\pm)}_{\ell n}}{\imag \omega^{\,(\pm)}_{\ell n}} \right]^{\tfrac{1}{2}} \omega^{\,(\pm)}_{\ell n} \, B^{\,(\pm)}_{\ell n},
\end{equation}
that we can compute with the numerical data obtained in the previous section.

In Fig.~\ref{fig:heff} we show the effective amplitudes of the fundamental quasinormal modes of polar (solid line)
and axial (dashed line) parities as functions of $\varepsilon$.
We see that the ``polar-wave amplitude'' dominates over its axial counterpart for $\varepsilon \lesssim 0.012$, above which the axial mode dominates.
This suggests that \emph{at least for some initial data} (see the ``asymptotic approximation'' of Ref.~\cite{Andersson:1996cm}) that the dominant contribution to the
gravitational-wave ringdown amplitude comes from the polar perturbations conditional, naturally, also on the cutoff $\varepsilon_{\rm max}$ for the onset on nonlinearities.

The dependence on this statement on the initial data (or source of perturbation) can be seen from source-dependent term in the quasinormal-mode excitation amplitude~\eqref{eq:exc_fac_source}. As an extreme example, it suffices to recall that
$I^{(-)}_n$ vanishes for a test particle radially infalling into a Schwarzschild black hole even if $B^{(-)}_n$ is nonzero.
Therefore, quasinormal modes of axial parity are not excited in this situation.
Nonetheless, we note that previous works in general relativity for plunging test particles~\cite{Silva:2023cer}
and in the close-limit approximation~\cite{LeTiec:2009yf} do find that the amplitude associated to perturbations
of polar parity to be larger than those of axial parity. We conjecture that this will also be the case in the EFT
studied here.

\section{Conclusions and outlook}
\label{sec:conclusions}
Motivated by a dearth of understanding about the quasinormal spectrum and about
how these quasinormal modes are excited in extensions to general relativity, we
reexamined and extended previous literature on perturbations of nonrotating black
holes in an EFT of general relativity.
Using a ``quick and dirty'' phase-amplitude method~\cite{Glampedakis:2003dn},
we computed both the first quasinormal-frequency overtones and the
quasinormal-mode excitation factors.
We found that the overtone frequencies (and their respective excitation
factors) are more sensitive than the fundamental quasinormal modes to the
lengthscale $l$ introduced by the higher-derivative operators
in the EFT.
We interpreted these results from an EFT perspective, and identified the domain
of validity of the EFT description; see Fig.~\ref{fig:spectrum_soft_uv}.
We also suggested the existence of an upper bound on the overtone number $n$,
above which a UV completion of the EFT becomes necessary to fully
characterize the black hole's quasinormal mode spectrum.
In addition, we presented a simple explanation for the inequivalence
between the spectra of quasinormal modes of polar and axial parities in the
EFT.

Let us put our findings in perspective.
In Sec.~\ref{sec:results} we argued that the overtones can be interpreted as high-frequency perturbations~\cite{Maggiore:2007nq},
and hence they probe the structure of the black-hole effective potential region near the horizon.
This is the region of spacetime where the EFT corrections are most significant;
this is unsurprising given that the EFT introduces higher powers of the curvature
that become relevant near the horizon.
Following this reasoning, it is not unreasonable to conjecture that this \emph{sensitivity of the
overtones to new lengthscales would also be seen in other theories that involve higher-curvature
corrections to general relativity, including those that introduce couplings to extra degrees of freedom.\footnote{While this paper was being completed,
a preprint presented the same reasoning~\cite{Hirano:2024fgp}.}}

It is also tempting to interpret our results in terms of an instability of the quasinormal-mode spectrum
induced by the higher-derivative operators in the EFT. In the language of Jaramillo et al.~\cite{Jaramillo:2020tuu}, this would correspond to an instability of the overtones; see e.g., Refs.~\cite{Aguirregabiria:1996zy,Nollert:1996rf,Nollert:1998ys} for earlier related works.
Confirmation of this interpretation would require an analysis of the quasinormal pseudospectrum associated to Eq.~\eqref{eq:eqs_rwz_fd} following, e.g., Ref.~\cite{Jaramillo:2020tuu}. If confirmed, it would suggest
that the \emph{instability of the overtones is a general expectation from an EFT perspective}.
It would then be interesting to understand how this ties with our conclusion that one cannot describe the quasinormal-mode spectrum past a certain overtone number without invoking the UV completion of the EFT.

Our calculation of the quasinormal-mode excitation factors and the observation that those associated to overtone are \emph{also} sensitive to the near-horizon modifications induced by the EFT has implications to the signatures of these corrections in gravitational-wave observations. We presented a first, but limited, analysis focusing on the implications of isospectrality breaking in Sec.~\ref{sec:results}
(see Fig.~\ref{fig:heff}), but more work is evidently needed~\cite{Silva:2024prep}.



\section*{Acknowledgements}
We thank Karim van Aelst, Alessandra Buonanno, Pablo A. Cano, Kyriakos Destounis, M. V. S. Saketh,
Jun Zhang, and Helvi Witek for discussions.
Some of our calculations were done with the {\sc Mathematica} packages
{\sc xPert}~\cite{Brizuela:2008ra} and {\sc Invar}~\cite{Martin-Garcia:2007bqa,Martin-Garcia:2008yei},
parts of the {\sc xAct/xTensor} suite~\cite{Mart_n_Garc_a_2008,xAct}.
This work makes use of the Black Hole Perturbation Toolkit~\cite{BHPToolkit}, in particular the \texttt{qnm} package~\cite{Stein:2019mop}.
H.O.S. acknowledges funding from the Deutsche Forschungsgemeinschaft
(DFG)~-~Project No.:~386119226.
K.G. acknowledges support from research grant No.~PID2020-1149GB-I00
of the Spanish Ministerio de Ciencia e Innovaci\'on.
K.Y. acknowledges support from NSF Grant No.~PHY-2207349, No.~PHY-2309066, No.~PHYS-2339969, and the Owens Family Foundation.

\appendix

\section{Derivation of the black hole solution}
\label{app:bh_derivation}

To obtain the black-hole spacetime described by the line element~\eqref{eq:line_element},
we solve the field equations~\eqref{eq:field_equations_schematic} perturbatively in $\varepsilon$.
To do so, we take $N$ and $f$ to be deformations away from their Schwarzschild expressions:
\begin{subequations}
\label{eq:ansatz_Nf}
\begin{align}
    f &= 1 - 2 \barm /r + \varepsilon \, \delta f,
    \\
    N &= 1 + \varepsilon \, \delta N,
\end{align}
\end{subequations}
where $\barm$ is a positive constant, $\varepsilon = \lambda \, l^{4} /
\bar{M}^4$, and $\delta f$ and $\delta N$ are functions of $r$. From the
$tt$ and $rr$ components of field equations, we find that $\delta N$ and
$\delta f$ obey the decoupled first-order differential equations:
\begin{subequations}
\begin{align}
    \left( \frac{\dd }{\dd r} + \frac{1}{r} \right) {\delta f} &=
    \frac{1080}{r}\frac{\barm^6}{r^6}
    - \frac{2352}{r} \frac{\barm^7}{r^7},
    \\
    \frac{\dd (\delta N)}{\dd r}  &= \frac{648}{r} \frac{\barm^6}{r^6}.
\end{align}
\end{subequations}
The solutions of these equations are
\begin{subequations}
\label{eq:deltafn_sols}
\begin{align}
    \delta f  &= \frac{c_1}{r}
    + \frac{216\barm^6}{r^6}
    - \frac{392\barm^7}{r^7},
    \\
    \delta N  &= c_{2} - \frac{108\barm^6}{r^6},
\end{align}
\end{subequations}
where $c_1$ and $c_2$ are integration constants. They can be fixed
by examining the far-field expansion of $g_{tt} = - N^2 f$,
\begin{equation}
    - g_{tt} \simeq 1 + 2 \varepsilon c_{2} - [\, 2\barm  - \varepsilon \, (c_{1} - 4 \barm c_{2})\,] / r.
\end{equation}
We can then set $c_{2} = 0$, as it amounts to a shift in the time coordinate $t$.
From the $r^{-1}$ term, we identify the ADM mass of the spacetime to be
\begin{equation}
    M = \barm - \varepsilon \, c_{1}/2.
    \label{eq:m_adm}
\end{equation}
Hence, the bare mass $\barm$ is renormalized by the dimension-six operators in the action.
We now solve Eq.~\eqref{eq:m_adm} for $\barm$, noticing that
$\varepsilon = \lambda \, l^4 / \barm^4 = \lambda \, l^4 / M^4$ to ${\cal O}(\varepsilon)$.
The result is $\barm = M + \varepsilon \, c_{1}/ 2$, which substituted together
with $c_{2} = 0$ in Eqs.~\eqref{eq:deltafn_sols} and~\eqref{eq:ansatz_Nf},
results in Eq.~\eqref{eq:metric_Nf} to ${\cal O}(\varepsilon)$.

\section{The Petrov classification of the black hole solution}
\label{app:petrov}

The Petrov type of a spacetime can be identified by constructing a null tetrad $l^\mu$, $n^\mu$, $m^\mu$ and its complex conjugate $\bar m^{\mu}$ and computing the Newman-Penrose scalars. The tetrad satisfies the normalization $l^\mu n_\mu = -1$, $m^\mu \bar m_\mu = 1$ with all the other contractions set to vanish. In this Appendix, we follow Refs.~\cite{Stephani:2003tm,Campanelli:2008dv} to determine the Petrov type of the
black hole solution given by Eqs.~\eqref{eq:line_element} and~\eqref{eq:resum_gtt_grr}.

When the spacetime is algebraically special (having at least one degenerate principal null direction), the following condition is satisfied
\begin{equation}
I^3 = 27 J^2,
\label{eq:IJ}
\end{equation}
where
\begin{align}
\label{I-eq}
I &= \frac{1}{2} \tilde{C}_{\alpha\beta\gamma\delta} \, \tilde{C}^{\alpha\beta\gamma\delta}, \nonumber \\
&=3 \Psi_2^2-4\Psi_1\Psi_3 + \Psi_4 \Psi_0,
\\ \nonumber \\
\label{J-eq}
J &=  -\frac{1}{6}  \tilde{C}_{\alpha\beta\gamma\delta} \, \tilde{C}^{\gamma \delta}{}_{\mu\nu} \, \tilde{C}^{\mu\nu\alpha\beta} , \nonumber \\
&= -\Psi_2^3 + 2 \Psi_1 \Psi_3 \Psi_2 + \Psi_0 \Psi_4 \Psi_2 - \Psi_4 \Psi_1^2 - \Psi_0 \Psi_3^2,
\end{align}
where we defined
\begin{equation}
\tilde{C}_{\alpha\beta\gamma\delta} = \frac{1}{4} \left( C_{\alpha\beta\gamma\delta} + \frac{i}{2}\epsilon_{\alpha\beta\mu\nu} C^{\mu\nu}{}_{\gamma\delta} \right),
\end{equation}
for a Weyl tensor $C_{\alpha\beta\gamma\delta}$, Levi-Civita tensor $\epsilon_{\alpha\beta\mu\nu}$, and where
$\Psi_i$ are the Newman-Penrose Weyl scalars with the only restriction of $\Psi_4 \neq 0$:
\allowdisplaybreaks
\begin{subequations} \label{eq:def_npw_scalars}
\begin{align}
\Psi_{0} &=C_{\alpha \beta \gamma \delta} \, l^{\alpha} m^{\beta} l^{\gamma} m^{\delta}, \\
\Psi_{1} &=C_{\alpha \beta \gamma \delta} \, l^{\alpha} n^{\beta} l^{\gamma} m^{\delta}, \\
\Psi_{2} &=C_{\alpha \beta \gamma \delta} \, l^{\alpha} m^{\beta} \bar{m}^{\gamma} n^{\delta}, \\
\Psi_{3} &=C_{\alpha \beta \gamma \delta} \, l^{\alpha} n^{\beta} \bar{m}^{\gamma} n^{\delta}, \\
\Psi_{4} &=C_{\alpha \beta \gamma \delta} \, n^{\alpha} \bar{m}^{\beta} n^{\gamma} \bar{m}^{\delta}.
\end{align}
\end{subequations}
In particular, $I$ and $J$ are nonvanishing for Petrov type D (and II).

To further determine the Petrov type, we study the following relations that are invariant under a tetrad rotation and hold for type D (and III):
\begin{equation}
K=0, \quad N-9L^2=0,
\label{conditionD}
\end{equation}
where
\allowdisplaybreaks
\begin{subequations}
\begin{align}
K & =  \Psi_1 \Psi_4^2 - 3 \Psi_4 \Psi_3 \Psi_2 + 2 \Psi_3^3, \\
\label{L}
L & =  \Psi_2 \Psi_4 - \Psi_3^2, \\
\label{N}
N & =  \Psi_4^3 \Psi_0 - 4 \Psi_4^2 \Psi_1 \Psi_3 + 6 \Psi_4 \Psi_2 \Psi_3^2 - 3 \Psi_3^4.
\end{align}
\end{subequations}
To summarize, the spacetime is type D if Eqs.~\eqref{eq:IJ} and~\eqref{conditionD} are satisfied for nonvanishing $I$ and $J$.

Let us now apply the above formulation to the black hole solution described by Eqs.~\eqref{eq:line_element} and~\eqref{eq:resum_gtt_grr}.
One of the simplest null tetrad is found as
\allowdisplaybreaks
\begin{widetext}
\begin{subequations} \label{eq:tetrads}
\begin{align}
l^\mu &= \left[\frac{r}{r-\rh}, 1
-\frac{5}{8} \frac{M}{r}
\varepsilon \left(
1+\frac{2 M}{r}+\frac{4 M^{2}}{r^{2}}+\frac{8 M^{3}}{r^{3}}+\frac{16 M^{4}}{r^{4}}
-\frac{704}{5} \frac{M^{5}}{r^{5}}
\right)
 ,0 ,0 \right] +\mathcal{O}(\varepsilon^2), \\
n^\mu &= \left[ {\frac {3\,r-\rh}{2(r-\rh)}}
+\frac{5}{16} \frac{M}{r}
\varepsilon\left(
1 +\frac{2 M}{r}+\frac{4 M^{2}}{r^{2}}+\frac{8 M^{3}}{r^{3}} +\frac{16 M^{4}}{r^{4}} +\frac{32 M^{5}}{r^{5}}\right), \right. \nonumber \\
&\left. \quad
{\frac {r+\rh}{2r}}
-\frac{5}{8} \frac{M}{r}
\varepsilon\left(
1+\frac{2 M}{r}+\frac{4 M^{2}}{r^{2}}+\frac{8 M^{3}}{r^{3}}+\frac{16 M^{4}}{r^{4}}
-\frac{272}{5} \frac{M^{5}}{r^{5}}
-\frac{864}{5} \frac{M^{6}}{r^{6}}
\right)
 , \frac{\sqrt{2}}{r} , 0 \right] +\mathcal{O}(\varepsilon^2), \\
m^\mu &= \left[\frac{r}{r-\rh} , 1
-\frac{5}{8} \frac{M}{r}
\varepsilon \left(
1+\frac{2 M}{r}+\frac{4 M^{2}}{r^{2}}+\frac{8 M^{3}}{r^{3}}+\frac{16 M^{4}}{r^{4}}
-\frac{704}{5} \frac{M^{5}}{r^{5}}
\right)
, \frac{1}{\sqrt{2}r} ,  \frac{\ii}{\sqrt{2}r \sin\theta} \right] +\mathcal{O}(\varepsilon^2).
\end{align}
\end{subequations}
\end{widetext}
Here, at $\mathcal{O}(\epsilon)$, we substituted $\rh = 2M$ to simplify the expressions.
We checked that the Petrov type of the black hole does not change even if we leave $\rh$ arbitrary.
Then, using Eqs.~\eqref{eq:def_npw_scalars} and~\eqref{eq:tetrads}, we find that $\Psi_i$ are given by
\allowdisplaybreaks
\begin{subequations}
\begin{align}
\Psi_0 &= \Psi_1 = \mathcal{O}(\varepsilon^2), \\
\Psi_2 &= {\frac {\rh}{2{r}^{3}}}+\frac{5M}{16r^3}\varepsilon \left(1-\frac{2304 M^{5}}{5 r^{5}} +\frac{960 M^{6}}{r^{6}}
 \right)+\mathcal{O}(\varepsilon^2), \\
\Psi_3 &= 3 \Psi_2\,, \quad \Psi_4 = 6 \Psi_2.
\end{align}
\end{subequations}

We can now obtain the Petrov type of the black hole. First,
the Weyl scalar invariants $I$ and $J$ are
given by
\begin{subequations}
\begin{align}
I &= {\frac {3{\rh}^{2}}{4{r}^{6}}}+\frac{15 M^2}{8r^6}\varepsilon\left(1-\frac{2304 M^{5}}{5 r^{5}}+\frac{960 M^{6}}{r^{6}} \right)+\mathcal{O}(\varepsilon^2), \\
J&= -{\frac {{\rh}^{3}}{8{r}^{9}}}-\frac{15 M^3}{16r^9}\varepsilon\left(1-\frac{2304 M^{5}}{5 r^{5}}+\frac{960 M^{6}}{r^{6}} \right)
+\mathcal{O}(\varepsilon^2),
\end{align}
\end{subequations}
which leads to
\begin{equation}
I^3 - 27 J^2 = \mathcal{O}(\varepsilon^2).
\end{equation}
Second, $K$, $L$ and $N$ are given by
\begin{subequations}
\begin{align}
K &= \mathcal{O}(\varepsilon^2), \\
L &= -{\frac {3{\rh}^{2}}{4{r}^{6}}}
-\frac{15 M^2}{8r^6}
\varepsilon\left(1-\frac{2304 M^{5}}{5 r^{5}}+\frac{960 M^{6}}{r^{6}} \right)
 + \mathcal{O}(\varepsilon^2), \\
N &= {\frac {81{\rh}^{4}}{16{r}^{12}}}+\frac{405 M^4}{4r^{12}}\varepsilon\left(1-\frac{2304 M^{5}}{5 r^{5}}+\frac{960 M^{6}}{r^{6}} \right)
 + \mathcal{O}(\varepsilon^2),
\end{align}
\end{subequations}
which leads to
\begin{equation}
N-9L^2 = \mathcal{O}(\varepsilon^2).
\end{equation}
Thus, to the order we worked on, $I$ and $J$ are nonvanishing and Eqs.~\eqref{eq:IJ} and~\eqref{conditionD} are satisfied, so the spacetime is of Petrov type D.

\section{Derivation of Eq.~\eqref{eq:resum_gtt_grr}}
\label{app:resummation}

Let us derive Eq.~\eqref{eq:resum_f} first. We start by adding and subtracting
$\rh / r$ to Eq.~\eqref{eq:grr_not_resum} and collect the terms as,
\begin{equation}
    f^{-1} =
    \left[
    1 - \frac{\rh}{r} + \left( \frac{\rh - 2M}{r} + \varepsilon \, \delta f \right)
    \right]^{-1}.
    \label{eq:grr_intm}
\end{equation}
From Eq.~\eqref{eq:event_horizon} we see that $\rh - 2M$ is of order $\varepsilon$
\begin{equation}
    \rh - 2M = \varepsilon \, \delta \rh = - \tfrac{5}{8} \, \varepsilon M.
    \label{eq:delta_rh}
\end{equation}
We can then rewrite Eq.~\eqref{eq:grr_intm}, by factoring out $1 - \rh / r$:
\begingroup
\begin{align}
f^{-1} &= \left(1-\frac{\rh}{r}\right)^{-1}
\, \left[
1 + \varepsilon \, \frac{{\delta \rh}/{r} + \delta f}{1 - \rh / r}
\right]^{-1},
\nn
&\simeq \left(1-\frac{\rh}{r}\right)^{-1}
\, \left[
1 - \varepsilon \, \frac{{\delta \rh}/{r} + \delta f}{1 - 2 M / r}
\right],
\end{align}
\endgroup
where we replaced $\rh / r$ by its ${\cal O}(\varepsilon^0)$ value in the second line.
We now use the explicit forms of $\delta \rh$ and $\delta f$, given in Eqs.~\eqref{eq:delta_f} and~\eqref{eq:delta_rh},
to find
\begin{align}
\label{eq:grr_resum_final}
f^{-1} &= \left(1-\frac{\rh}{r}\right)^{-1}  \, \left[
1 - \varepsilon \left( 1 - \frac{2M}{r} \right)^{-1}
\left( - \frac{5}{8}\frac{M}{r}
\right.
\right.
\nn
& \left.\left. \quad
+ \, \frac{216 M^6}{r^6} - \frac{392 M^7}{r^7}
\right)\right].
\end{align}
Although not evident, the term proportional to $\varepsilon$ inside the square brackets is regular at $r = 2 M$,
with value $-3/2$. To see this explicitly, we use the factorization:
\begin{align}
&- \frac{5}{8} \frac{M}{r} + \frac{216M^6}{r^6} -  \frac{392M^7}{r^7} =
- \left(1-\frac{2M}{r}\right)
\left( \frac{5}{8} \frac{M}{r}
\right.
\nn
&\left.
+ \frac{5}{4}\frac{M^2}{r^2} + \frac{5}{2}\frac{M^3}{r^3} + \frac{5 M^4}{r^4} + \frac{10 M^5}{r^5} - \frac{196 M^6}{r^6}
\right).
\end{align}
Using this result, we obtain Eq.~\eqref{eq:resum_f}.
Equation~\eqref{eq:gtt_resum_final} is derived in the same manner. We find
\begin{align}
    N^2 f &= \left(1 - \frac{\rh}{r}\right) \,
    \left[
    1 + 2 \varepsilon \, \delta N + \varepsilon \, \frac{\delta \rh/r + \delta f}{1-2M/r}
    \right],
    \nonumber \\
    &= \left(1 - \frac{\rh}{r}\right)\left[ 1 - \varepsilon \, \left(
    \frac{5}{8}\frac{M}{r} + \frac{5}{4}\frac{M^2}{r^2} + \frac{5}{2}\frac{M^3}{r^3} + \frac{5 M^4}{r^4}
    \right. \right.
    \nonumber \\
    &\quad \left.\left. +\frac{10 M^5}{r^5}+\frac{20 M^6}{r^6}\right)\right],
\end{align}
where we used $\delta N = - 108 \,  {M^6} / {r^6}$.
%

\section{The tortoise coordinate}
\label{app:tortoise}

In this appendix, we analyze in detail the properties of the tortoise
coordinate $x$, defined in Eq.~\eqref{eq:def_tortoise},
\begin{equation*}
\dd x / \dd r = 1 / (Nf).
\end{equation*}
We use the resummation recipe introduced in Sec.~\ref{sec:bh_solution}
and detailed in Appendix~\ref{app:resummation} to
rewrite Eq.~\eqref{eq:def_tortoise} as
\begin{align}
\frac{\dd x}{\dd r} &= \left(1 - \frac{\rh}{r}\right)^{-1} \,
\left[ 1 + \varepsilon \left(
 \frac{5}{8}\mr{}
+ \frac{5}{4}\mr{2}
+ \frac{5}{52}\mr{3}
\right. \right.
\nn
&\quad \left. \left.
+ \frac{5 M^4}{r^4}
+ \frac{10 M^5}{r^5}
- \frac{88 M^6}{r^6}
\right) \right].
\end{align}
This differential equation can be solved analytically, and the solution
can be schematically written as
\begin{align}
x = r + \rh \, \log(r/\rh - 1) + \varepsilon \, \delta x(r),
\label{eq:x_analytical}
\end{align}
where we set the integration constant to be $-\rh \log \rh$. The expression for $\delta x$ is somewhat lengthy and
we omit it for brevity. In the limit of general relativity ($\varepsilon = 0$),
we recover the usual Schwarzschild formula
\begin{equation}
x = r + 2M \log[r/(2M) - 1].
\end{equation}

For $x$ to be a bona fide tortoise
coordinate, we expect that $x \to -\infty$ as $r \to \rh$
and that $x \to \infty$ as $r \to \infty$. Whether this is the case for all values of $\varepsilon$
is not immediately evident. Let us first consider the limit of spatial infinity.
In this limit, an expansion of Eq.~\eqref{eq:x_analytical} yields:
\begin{equation}
x \simeq r + \rh \left( 1 + \varepsilon \, \frac{5}{8} \frac{M}{\rh} \right) \log r,
\quad \textrm{for} \quad  r/\rh \gg 1.
\end{equation}
Because the term in parentheses is ${\cal O}(1)$, and because $r$ grows faster
than $\log r$, we conclude that $x \to \infty$ when $r \to \infty$, as desired.

We now consider the near-horizon limit. In this limit, an expansion of Eq.~\eqref{eq:x_analytical} yields:
\begin{align}
x &\simeq \rh + \rh \log(r - \rh)
+ \varepsilon \, \rh \, [\, p_{0} + p_{1} \log \rh
\nn
&\quad + p_{2} \log(r - \rh) \, ] ,
\quad \textrm{for} \quad  r/\rh \sim 1,
\label{eq:x_near_horizon}
\end{align}
where $p_{i}$ ($i = 1$,~$2$,~$3$) are sextic polynomials in $M/r$. The coefficients
in the polynomials are not all positive, and, consequently, we need to look whether
$x \to -\infty$ as $r \to \rh$ in more detail.
We first note that the dominant terms in Eq.~\eqref{eq:x_near_horizon} for
$r \approx \rh$ are those proportional to $\log(r - \rh)$, i.e.,
\begin{align}
x &\simeq \rh (1 + \varepsilon \, p_{2} ) \log(r - \rh),
\label{eq:x_near_horizon_dominant}
\end{align}
where $p_2$ is:
\begin{equation}
p_2 =
\frac{5}{8}\frac{M}{\rh}
+ \frac{5}{4} \frac{M^2}{\rh^2}
+ \frac{5}{2} \frac{M^3}{\rh^3}
+ \frac{5 M^4}{\rh^4}
+ \frac{10 M^5}{\rh^5}
-\frac{88 M^6}{\rh^6}.
\end{equation}
In units in which $M = 1$ and for $\rh \approx 2$
[cf.~Eq.~\eqref{eq:event_horizon}], $p_2$ has a magnitude of ${\cal O}(10^{-1})$. Hence, depending
on the value of $\varepsilon$, $x$ can approach either $\pm \infty$ in the limit
$r \to \rh$.
Numerically, we found that $1 + \varepsilon \, p_2$  becomes negative for $\varepsilon \gtrsim 0.59$. This value of $\varepsilon$ is one order of magnitude larger than the
values we considered in the main text.
Hence, $x$, as given by Eq.~\eqref{eq:x_analytical},
has the desired properties of a tortoise coordinate for all practical purposes.

\section{Coefficients in the effective potential}
\label{app:effective_potential_coeffs}

In this appendix, we present the coefficients $v_{i \ell}^{\,(\pm)}$ that appear in the
EFT corrections $\delta V_{\ell}^{\,(\pm)}$
to the Zerilli and Regge-Wheeler potentials~\eqref{eq:effective_potentials_eft_schematic}.
The coefficients $v_{i \ell}^{\ps}$ in the polar-parity potential are
\begingroup
\allowdisplaybreaks
\begin{align}
    \label{eq:v_even}
    v^{\ps}_{1 \ell}  &= -5 \lambda_{\ell} ^2 \, (\lambda_{\ell} +1),
    \nn
    v^{\ps}_{2 \ell}  &= -5 \lambda_{\ell} ^2 \, (2 \lambda_{\ell} +5),
    \nn
    v^{\ps}_{3 \ell}  &= -5 \lambda_{\ell} \, (4 \lambda_{\ell} ^2+10 \lambda_{\ell} +9),
    \nn
    v^{\ps}_{4 \ell}  &= -5 \, (8 \lambda_{\ell} ^3+20 \lambda_{\ell} ^2+18 \lambda_{\ell} +9),
    \nn
    v^{\ps}_{5 \ell}  &= 10 \, [-8 \lambda_{\ell} ^3-20 \lambda_{\ell}^2-18 \lambda_{\ell}
    \nn
                      &\quad - {288 \lambda_{\ell}^3 \, (\ell ^2+\ell -6)}/{\Lambda_{\ell} }-9],
    \nn
    v^{\ps}_{6 \ell}  &= 4 \, \{
    176 \lambda_{\ell} ^3+116 \lambda_{\ell} ^2-90 \lambda_{\ell}
    \nn
                 &\quad + {54 \lambda_{\ell}^2 [ 15 \ell ^2 (\ell +1)^2-336 \ell  (\ell +1)+836]}/{\Lambda_{\ell}} - 45
    \},
    \nn
    v^{\ps}_{7 \ell}  &= 24 \, \{
        88 \lambda_{\ell} ^2-30 \lambda_{\ell}
        + 15 \lambda_{\ell}  \, [ 147\,  \ell ^2 (\ell +1)^2
                 \nn &\quad
        - 1304 \, \ell  (\ell +1)+2164 ]/\Lambda_{\ell} - 15
    \},
    \nn
    v^{\ps}_{8 \ell}  &= 144 \, \{ -5 + 44 \lambda_{\ell} + 3 \lambda_{\ell} \, [1073 \, \ell (\ell +1)-3988] / \Lambda_{\ell} \},
    \nn
    v^{\ps}_{9 \ell}  &= 6336 + [778608 \, \ell  (\ell +1)-1938240]/\Lambda_{\ell},
    \nn
    v^{\ps}_{10 \ell} &= 879552/\Lambda_{\ell}.
\end{align}
\endgroup
The coefficients $v_{i \ell}^{\mn}$ in the axial-parity potential are
\begingroup
\allowdisplaybreaks
\begin{align}
    \label{eq:v_odd}
    v^{\mn}_{1 \ell} &= -\tfrac{5}{8} \, \ell (\ell+1),
    \nn
    v^{\mn}_{2 \ell} &= -\tfrac{5}{4} \, (\ell^2+\ell-3 ),
    \nn
    v^{\mn}_{3 \ell} &= -\tfrac{5}{2} \, (\ell^2+\ell-3 ),
    \nn
    v^{\mn}_{4 \ell} &= -5 \, (\ell^2+\ell-3 ),
    \nn
    v^{\mn}_{5 \ell} &= 1430 \, \ell(\ell+1) - 8610,
    \nn
    v^{\mn}_{6 \ell} &= 41460 - {3332} \, \ell(\ell+1),
    \nn
    v^{\mn}_{7} &= -{48192}.
\end{align}
\endgroup
We recall that $\lambda_{\ell}$ and $\Lambda_{\ell}$, appearing in Eqs.~\eqref{eq:v_even}
and~\eqref{eq:v_odd}, are defined in Eq.~\eqref{eq:def_lambdas}.

\bibliography{biblio}
\end{document}